\documentclass{JHEP3}
\usepackage{amsmath,amsfonts,bbm,euscript}
\usepackage{cite}
\usepackage{epsfig}
\usepackage{graphicx}
\usepackage{psfrag}
\def\be{\begin{equation}}
\def\ee{\end{equation}}
\def\baray{\begin{eqnarray}}
\def\earay{\end{eqnarray}}
\def\ba{\begin{eqnarray}}
\def\ea{\end{eqnarray}}

\def\F{\cal{F}}
\newcommand{\rb}{\bar{\rho}}

\newcommand{\Sp}{S^{\prime}}

\title{Spontaneous Creation of Inflationary Universes and the Cosmic Landscape}
\author{Hassan Firouzjahi, Saswat Sarangi and S.-H. Henry Tye \\
 Laboratory for Elementary Particle Physics, Cornell
University, Ithaca, NY 14853 \\
E-mail: \email{firouzh@lepp.cornell.edu},
\email{sash@lepp.cornell.edu},
\email{tye@lepp.cornell.edu}}

\abstract{
We study some gravitational instanton solutions that offer 
a natural realization of the spontaneous creation of inflationary 
universes in the brane world context in string theory. 
Decoherence due to couplings of higher (perturbative) modes of 
the metric as 
well as matter fields modifies the Hartle-Hawking wavefunction 
for de Sitter space. Generalizing this new wavefunction to be 
used in string theory, we propose a principle in string 
theory that hopefully will lead us to the 
particular vacuum we live in, thus avoiding the anthropic principle.
As an illustration of this idea, we give a phenomenological 
analysis of the probability of quantum tunneling to various stringy 
vacua. We find that the preferred tunneling is to an inflationary 
universe (like our early universe), not to
a universe with a very small cosmological constant 
(i.e., like today's universe) and not to a 10-dimensional 
(or a higher dimensional supercritical) 
uncompactified de Sitter universe. Some solutions 
are interesting as they offer a cosmological mechanism for the 
stabilization of extra dimensions during the inflationary epoch.
}


\begin{document}

\section{Introduction}

String theory is the only known candidate for a unified theory of 
fundamental physics. Recent understandings in string theory and its 
compactifications \cite{Giddings:2001yu,Kachru:2003aw,
Saltman:2004sn,Denef:2004dm,Denef:2004ze}
have led to the realization that string theory has many vacua 
(here we include metastable vacua with lifetimes comparable to or 
larger than the age of our universe). The number of such vacua,
if not infinite, may be as huge as $10^{100}$ or larger 
\cite{Douglas:2003um,Ashok:2003gk,Douglas:2004kp,Weinberg:2000qm,
Susskind:2003kw,Susskind:2004uv,Banks:2003es}. 
How we end up in the particular vacuum we are in, that is, the 
particular site in this vast cosmic landscape, is a very important but 
highly non-trivial question.  
One may simply give up on this question by invoking the 
``anthropic principle''. More positively, one may take the optimistic 
view that there exists a principle which tells 
us why we end up in the particular string vacuum we are in. 
In this paper, we propose such a principle.
A better understanding of string theory and gravity may eventually 
allow us to check the validity of the idea, or maybe to improve on it.
In the meantime, we may use this proposal as a working hypothesis and 
a phenomenological tool. As a minimum, even if our specific proposal
turns out to be not quite correct, we hope it convinces some readers 
that such a 
principle does exist, allowing us to bypass the ``anthropic principle''. 

Since observational evidence \cite{Peiris:2003ff} of an inflationary 
epoch \cite{Guth:1981zm,Linde:1982mu,Albrecht:1982wi} is very strong,
we suggest that the selection of our particular vacuum state 
follows from the evolution of the inflationary epoch.
That is, our particular vacuum site in the cosmic landscape
must be at the end of a road that an 
inflationary universe will naturally follow. Any vacuum state that
cannot be reached by (or connected to) an inflationary stage can be 
ignored in the search of candidate vacua. That is, the issue of the
selection of our vacuum state becomes the question on the selection 
of an inflationary universe, or the selection of an original universe 
that eventually evolves to an inflationary universe, which then 
evolves to our universe today. Let us call this the 
``Selection of the Original Universe Principle'' or SOUP for short.
The landscape of inflationary states/universes should be much better 
under control, since the inflationary scale is rather close to
the string scale.
Here we propose that, by analyzing all known string vacua 
and string inflationary scenarios, one may 
be able to phenomenologically pin down SOUP, or at least 
discover some properties of such a principle, which may 
then help in the derivation of SOUP in string theory.
The key tool we shall use here is a modification of the Hartle-Hawking 
wavefunction \cite{Hartle:1983ai}.

Even if we understand inflation completely, its density perturbations, 
and all aspects of astrophysics related to galaxy and star formations,
no one should expect us to be able to calculate from first principle
the masses of our sun
and our earth and why our moon has its particular mass. On the other
hand, we do understand why the mass of our sun is 
not much bigger/smaller than its measured value. That is, 
our planetary system is a typical system that is expected based on
our present knowledge. As theorists, we are comfortable with this situation.
Along this line of thinking, one should feel content if the SOUP can 
show that our universe is among a generic set of preferred vacua, even 
if one fails to show why we must inevitably end in the precise vacuum 
state we are in. It is along this outwardly less ambitious, but 
probably ultimately more scientifically justified, direction that we 
are trying to reach. 

Modern cosmology aims to describe how our universe has evolved to 
its present state from a certain initial state. 
An appealing scenario, due to 
Vilenkin \cite{Vilenkin:1982de,Vilenkin:1983xq}, 
Hartle and Hawking \cite{Hartle:1983ai} and 
others \cite{Tryon:1973}, proposes that the inflationary universe 
was created by the quantum tunneling from ``nothing'',
that is, a state of no classical space-time. 
This quantum tunneling from nothing, or the ultimate free lunch, 
should be dictated by the laws of physics, and it avoids the 
singularity problem that would have appeared if one naively 
extrapolates the big bang epoch backwards in time. 
It is natural to extend this approach to the brane world scenario, 
to see how brane inflation \cite{Dvali:1998pa,
Kachru:2003sx,Firouzjahi:2003zy,Hsu:2003cy,Burgess:2004kv,Dasgupta:2004dw},
natural in superstring theory, may emerge. 
Our understanding of superstring theory has advanced considerably in
recent years so that it is meaningful to address this question.
In this paper, we would like to consider some simple models motivated by 
superstring theory, to see what new features and issues may arise. 
We consider this as a first step towards the goal of understanding what 
string theory is trying to tell us about the origin of our Universe.     
\begin{figure}
\begin{center}
\epsfig{file=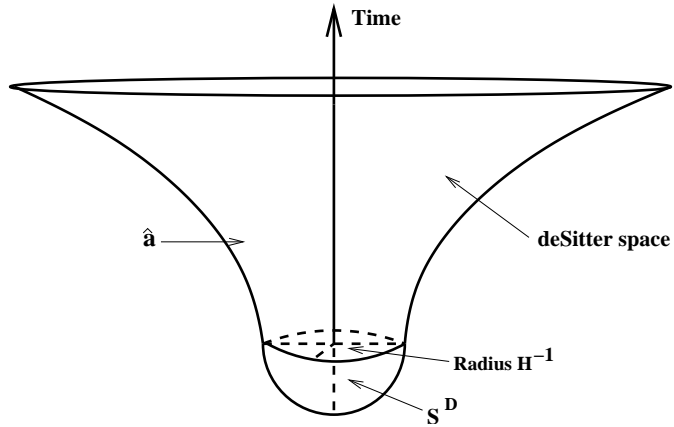, width=9cm}
\vspace{0.1in}
\caption{Starting with nothing, quantum tunneling happens via a 
$S^{D}$ instanton with radius $a=1/H$ to a $D$-dimensional de Sitter 
universe, which then grows to $\hat a$ and beyond.
}
\label{fig1}
\end{center}
\end{figure}

\begin{figure}
\begin{center}
\epsfig{file=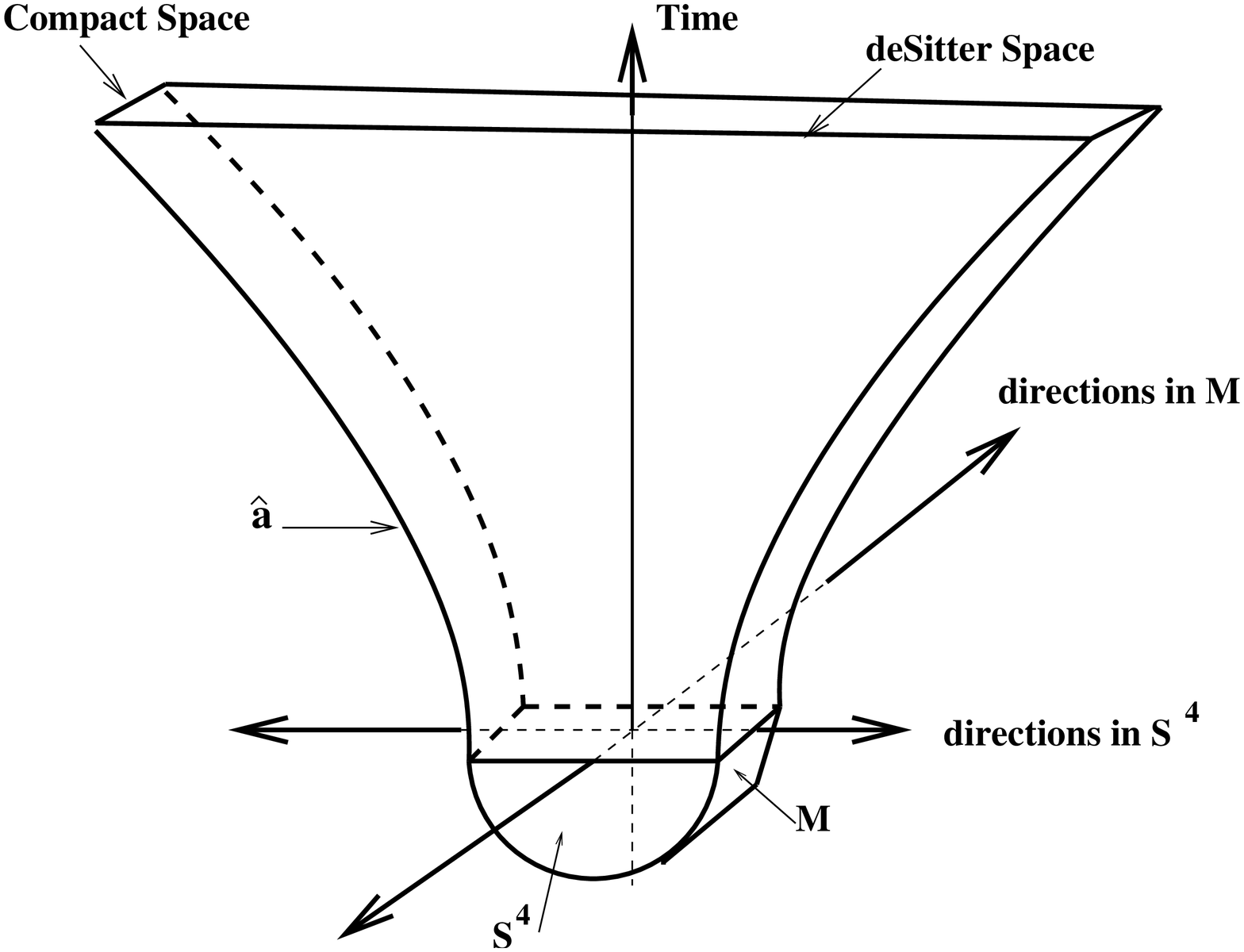, width=9cm}
\vspace{0.1in}
\caption{A $S^4 \times M$ instanton tunneling to the 
4-dimensional de Sitter universe with a cosmologically stable 6 spatial 
dimensional space $M$. Some examples are $M=S^6, 
S^2 \times S^2 \times S^2$, three-fold Calabi-Yau manifold with 
fluxes etc.
}
\label{fig2}
\end{center}
\end{figure}
For SOUP in this free lunch, we need to determine the relative 
probability amplitude of tunneling from nothing to a variety of
universes. This then allows us to select the universe with the 
largest probability amplitude.
Suppose that the probability amplitude of this tunneling to a de Sitter 
universe is given by the Hartle-Hawking wavefunction of the universe 
with a cosmological constant $\Lambda$, 
in terms of the Euclidean action $S_E$ as 
\ba
\label{HHWF1}
\Psi_{HH} \sim e^{-S_E} = e^{3\pi/2 G_N \Lambda} 
\ea
Note that $S_E$ is unbounded from below \cite{Gibbons:1978ac} and
small $\Lambda$ is exponentially preferred.
Assuming the dark energy observed today is due to a very small
cosmological constant $\Lambda_{today}$, 
then it is exponentially more likely to 
tunnel directly to today's universe
(or our universe many billions years in the future) than to an 
inflationary universe 13.7 billions years ago. 
If there exists a solution with an even smaller 
cosmological constant, then that universe will be exponentially more
preferred.
Clearly this is worse than the naive anthropic principle, and the above 
formula must be modified. In fact, it was pointed out that the above 
probability amplitude is unstable to corrections \cite{Fischler:1990se}. 
Intuitively, tunneling directly to today's universe with its immense 
size ($a \sim \Lambda_{today}^{-1/2}$) must be suppressed. 

To tunnel from nothing to an inflationary universe that describes our 
early universe, we need a reason that selects a tunneling to some 
intermediate value of $\Lambda_{today} << \Lambda << G_N^{-1}$.
(The observational data suggests that the tunneling to an approximate 
de Sitter universe with a $\Lambda$ relatively close to the GUT scale 
is preferred.) 
As a phenomenological ansatz, we need to find an improved wavefunction 
by modifying the Hartle-Hawking wavefunction. Physically, we see 
that the above $\Psi_{HH}$ has not included the effects of matter fields
and the gravitational perturbative modes around the de Sitter metric.
Naively, one may think that those effects are small. Here we shall 
argue otherwise. 
  
We shall present 3 different (though ultimately equivalent) arguments 
for the necessity of such a modification: 
(i) destructive interference due to small fluctuations of large phases, 
(ii) quantum decoherence and (iii)
space-like brane in string theory. The first argument is mostly intuitive, 
while the last two suggest the specific way the wavefunction should 
be improved. 

Our first argument goes as follows: 
(1) In tunneling to a string vacuum state which then evolves 
classically involves first Euclidean and then Lorentzian time.
Treating time as a coordinate, this means complex metric 
(more precisely, complex lapse function) is involved. This 
implies that, in the sum over paths in the evaluation of the
wavefunction, we should include paths with complex metric, not 
just real metric. This was suggested by Halliwell and 
Louko \cite{Halliwell:1989vu} and others for 4-dimensional gravity.
(2) In summing over paths, the steepest descent method is employed.
This is standard practice. Here, we note that there is a large 
degeneracy in paths. That is, very different paths yields the 
same action with an imaginary part, $S=S_R +i S_I$. 
We expect this degeneracy to be lifted by the presence of gravitational
and matter modes interacting with the classical metric.
(3) When the phase $S_I$ is very
large (i.e., exponentially large compared to $\pi$) and $S$ fluctuates, 
the path degeneracy is lifted and the sum over paths will 
in general lead to destructive interference, so that quantum effects
(quantum tunneling here) will be suppressed. This is analogous to
the situation of a macroscopic particle (say a marble or a billiard ball) 
in quantum mechanics. Here we make the assertion that this decoherence 
takes place at the lifting of the above path degeneracy with very 
large $S_I$s. We find that a large $S_I$ phase
appears generically in the quantum tunneling to a vacuum state with

(a) a large inflating volume, i.e., de Sitter size,

(b) a small cosmological constant $\Lambda$, and/or 

(c) a large compactified extra dimensions.

\noindent (4) In particular, individual paths in the 
4-dimensional de Sitter case contribute to $\Psi$ an 
imaginary phase $\sim 1/G_N\Lambda$. So we expect destructive
interference to suppress the tunneling probability in any of 
these situations. In particular, 
the resulting effect following from the sum over paths 
will suppress the tunneling to a small $\Lambda$ universe. This 
result is very different from that suggested by Eq.(\ref{HHWF1}).
However, a direct sum over such paths may be difficult. Usually, a 
rotation to Euclidean space allows one to sum over all paths and
evaluate such an effect. This will lead to a new term in the 
wavefunction. So we are led to propose the following modification 
to the Hartle-Hawking wavefunction;
\be
\label{master}
\Psi \sim e^{\F}, \quad \quad {\F} = -S_E - {\cal{D}}
\ee
where the decoherence term ${\cal{D}}$ is real positive. 
This wavefunction reduces to the Hartle-Hawking wavefunction 
for ${\cal{D}}=0$. For SOUP, ${\cal{D}}$ should be large in 
any of the three situations listed above; 
that is, the larger is the universe or the smaller is the 
cosmological constant, the larger is the value of ${\cal{D}}$, 
so their tunneling is suppressed. 

The above argument can be made more precise in quantum 
decoherence. In decoherence, the classical metric (say, the 
cosmic scale factor $a$) is treated as 
the configuration variable while the perturbative modes around 
this metric and matter fields that couple to it are treated 
as the environment. The presence of the environment causes the 
quantum system to experience a dissipative dynamics, and the 
loss of quantum coherence results in the modification of the 
Hartle-Hawking wavefunction. In fact, the form of the leading 
term in ${\cal{D}}$ can be easily found. 
A simple generalization of the known results
\cite{Kiefer:1987ft,Halliwell:1989vw,Kiefer:1989ud} (which were 
used to justify the classical treatment of time) to the tunneling 
case in string theory yields
\ba
\label{DV_9}
{\cal{D}} = c V = c \left( \frac{M_s}{2 \pi} \right)^9 V_9 +  ...
\ea
Here, $c$ is a dimensionless constant and 
$M_s=1/\sqrt{\alpha^{\prime}}$ is the 
superstring scale. Here $V$ is the dimensionless ``spatial volume'' 
(measured in $l_s \equiv 2\pi/M_s$) of the de Sitter (or any other) 
instanton, or the ``area of the boundary'' towards the end of 
tunneling.
This is crudely the transition region from
Euclidean to Lorentzian space (see Figures \ref{fig1} and \ref{fig2})). 
The origin of this term may be argued on physical grounds.
Each mode of the environment supplies a mode-independent 
suppression factor so the resulting suppression factor is 
proportional to the total number of modes. Modes with wavelength 
longer than some fixed scale are unobservable 
and so should be traced over in the density matrix to yield the 
reduced density matrix. This cut-off implies that the total number 
of modes traced over is proportional to $V_9$, thus yielding the 
above ${\cal{D}}$ term. 
A constant term in $\cal{D}$ may be absorbed into 
the normalization of $\Psi$. Quantum contributions of matter fields 
may also contribute to the prefactor ${\cal P}$ in 
$\Psi = {\cal P} \exp ({\F})$.

Our third argument is indirect. 
In principle, ${\cal{D}}$ should be calculable in string theory, 
that is, the ${\cal{D}}$ term must have a calculable value for 
each of the potential string vacua, stable, metastable, or unstable. 
We may describe the tunneling as due to the presence of a 
$S$-brane. Since we are dealing with complex time (or Euclidean to 
Lorentzian time), we call such a space-like brane a $\Sp$-brane.
A boundary term involving a $\Sp$-brane is proportional to $V_9$,
leading us to the same term in ${\cal{D}}$ (\ref{DV_9}).

The $V_9$ term suppresses universes with a small 
$\Lambda$ and/or a large inflationary size. 
To suppress universes with very large (or uncompactified)
extra dimensions, we need additional terms. 
(We do not expect ${\cal D}$ to be simple.)
Possible simple terms are $V_{10}^2$, $V_9 V_{10}$ etc. 
In cases where the extra dimensions are dynamically stabilized
(even only to a metastable state), we may use the ${c}V$ 
term only (or a $V_{10}$ term) to illustrate the approach. 
$c$ may be a function of the vacuum state as well.
It is important to determine the
functional form of ${\cal{D}}$ in string theory in a 
more careful analysis. 
For our purpose here, the above choice of ${\cal D}$ is enough to 
explain our main points. 

In string theory, both
$S_E$ and ${\cal{D}}$ (and so $\F$ and $\Psi$) should be calculable 
for each possible vacuum state.
The inflationary vacuum state that 
ends in our today's universe should be the vacuum state that 
maximizes $\Psi$ or $\F$. If this SOUP program is successful, we 
should be able to understand why we are where we are (why 3 large 
dimensions, why at most $N=1$ supersymmetry, why 
$SU(3) \times SU(2) \times U(1)$ etc.), thus avoiding the 
anthropic principle. 

As an illustration, we simply consider phenomenologically 
the above explicit term in ${\cal{D}}$ that will provide 
the suppressions we are looking for and can be applied to 
known string vacua.
To illustrate the SOUP idea, let us apply the above specific 
ansatz to the brane inflationary scenario proposed by 
KKLMMT \cite{Kachru:2003sx}.
We use $G_N$ and the COBE density perturbation data to crudely fix
$M_s$, $\Lambda$ and $c$, namely $M_s \simeq 4 \times 10^{17}$ GeV,
$\sqrt{\Lambda} \sim 10^{14}$ GeV and 
${c} \simeq 10^{-3}$. 
Maximizing $\F$ fixes the fluxes that stabilize 
the Calabi-Yau manifold.
We find that ${\F} \simeq 10^{18}$. Using the determined value of 
${c}$, we find the value of 
$\F$ for the tunneling directly to a KKLT vacuum \cite{Kachru:2003aw}
without inflation (which has ${\F} <0$).
On the other hand, tunneling to a ten-dimensional de Sitter universe
$S^{10}$, or other similar universes such as $S^4 \times S^6$, 
$S^5 \times S^5$ etc. (with $\Lambda$ suppied by 
$D9-{\bar D}9$-brane pair)
has ${\F} \simeq 10^{9}$, largely independent of the value of $c$.
In summary, the inflationary universe is much preferred among the vacua 
we examined.
This is not surprising, since the functional form of $\F$ 
for a $D$-dimensional inflationary universe (with the remaining 
compactified dimensions stabilized) is,
\be
{\F} = \frac{a}{\Lambda^{(D-2)/2}} -\frac{b}{\Lambda^{(D-1)/2}} + ... 
\ee
\begin{figure}
\hspace{4cm}${\F}$
\begin{center}
\epsfig{ file=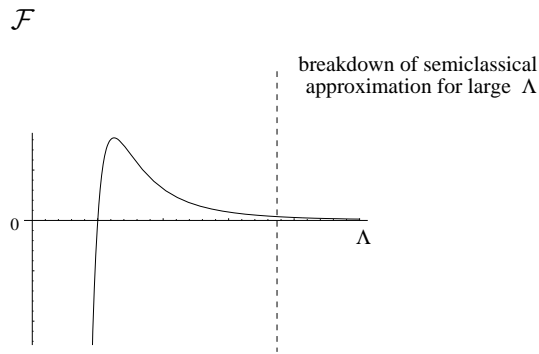, width=7cm} 
\vspace{0.1in}
\caption{${\F}$ as a function of $\Lambda$. The semi-classical approximation 
breaks down (indicated by the dashed vertical line at the right) when 
$\Lambda$ is large.
}
\label{fig3}
\end{center}
\end{figure}
As shown in Figure \ref{fig3}, for generic constants $a$ and $b$,
an intermediate value of $\Lambda$ is clearly preferred over a 
very large or a very small $\Lambda$. In particular, tunneling 
directly to a supersymmetric vacuum is severely suppressed. 
To see which inflationary scenario or some other vacuum state is
preferred via tunneling, a much more precise determination of $\Psi$
is required. Note that SOUP will not work
if we use the alternative to the Hartle-Hawking wavefunction 
\cite{Linde:1984mx,Vilenkin:1985dy}, where the sign of the first 
term becomes negative. In this case, large $\Lambda$ (easily achieved
with many brane-antibrane pairs) is preferred, which leads to 
the breakdown of the WKB approximation.

Although the details of this preliminary analysis is 
admittedly quite simplistic,
it does open the possibility that a good understanding of the 
decoherence term may select a particular inflationary state
which will then evolve to our today's universe after inflation.
A more detailed analysis may allow us to determine more completely
the functional form of ${\cal D}$ (and the value of $c$).
At the same time, as one examines more string theory vacua and 
inflationary solutions, one should be able to phenomenologically 
refine the functional form of ${\cal D}$ as well.
This is the basic point of SOUP. The reader may think that
SOUP is too good to believe. We argue that the anthropic 
principle is too hard to believe; almost any scientifically motivated
alternative is more desirable.

Slow-roll inflation usually implies some form of eternal inflation
\cite{Steinhardt:1982kg,Linde:1986fd}.
With eternal inflation, it is argued that the origin of the universe
issue is less pressing. Following our view point, this does not 
explain why the particular eternal inflationary universe is selected
in the first place. 
If eternal inflation has happened eternally, the anthropic principle 
must be invoked in this scenario to explain why the particular 
inflationary universe is selected. On the other hand,
SOUP will select the particular inflationary universe,
even if it has the eternal inflationary property.
A somewhat similar comment may be applied to the appearance of 
inflationary (or some other) universes from a ``meta-universe'' 
\cite{Dyson:2002pf,Albrecht:2004ke}. We must first understand
the origin of such a meta-universe.

If successful, this program may also be used to make predictions 
that can be tested in the near future. In the
brane inflationary scenario, cosmic strings are produced towards 
the end of inflation and they will leave signatures to be detected
\cite{Jones:2002cv,Sarangi:2002yt,Jones:2003da,Pogosian:2003mz,Copeland:2003bj,Dvali:2003zj,Leblond:2004uc,Jackson:2004zg}.
However, the type/property of the cosmic strings produced depends 
crucially on the particular brane inflationary scenario that had taken 
place. If SOUP can select the specific brane inflationary universe 
emerging from tunneling, it will give predictions on the type/property
of the cosmic strings to be detected.

It was pointed out that supercritical ($D>10$) string vacua 
\cite{Chamseddine:1992qu,Silverstein:2001xn,Maloney:2002rr}
may also provide a realistic description of today's universe.
However, supersymmetry is typically absent in such supercritical vacua, 
so low energy supersymmetry will not be detected in LHC and similar
experiments. It is fair to ask if SOUP can distinguish between 
critical and supercritical string vacua. Although SOUP 
clearly prefers $S^{10}$ over $S^D$ for $D>10$, with 
${\F}(D>10)<10^5$, 
it is not so clear for inflationary scenarios in critical versus 
supercritical string vacua, assuming viable supercritical 
inflationary scenarios can be realized. 
If so, the determination of the string coupling $g_s$ among various 
inflationary scenario via the maximization of $\F$ may be able to 
distinguish them. 
Clearly a better understanding of $\F$ should be valuable as well.
 
In this speculative paper, the following issues are discussed. 
In Sec. 2, we argue for the destructive interference in the path sum
and propose the wavefunction that includes the 
decoherence term. We discuss the decoherence approach 
to find that the form of the leading term in ${\cal D}$ is 
the above $V_9$ term.   
In Sec. 3, we apply this wavefunction for the quantum tunneling from 
nothing via a 10-dimensional de Sitter instanton $S^{10}$
to a 10-dimensional de Sitter space. 
The cosmological constant comes from the pair creation of branes, 
the simplest being $D$9-$\bar D$9-brane pairs in Type IIB models or 
non-BPS $D$9-branes in Type IIA models. In this paper, we shall 
focus on Type IIB or F theory vacua.  
We also consider other cases, such as $S^4 \times S^6$, $S^5 \times S^5$,
$S^4 \times S^2 \times S^2 \times S^2$
$S^3 \times S^3 \times S^4 $, $S^4 \times T^1S^2 \times T^3$ etc..
Quantum tunneling happens at a scale below the superstring scale, 
so that semi-classical gravitational analysis should be valid. 
Among this set of vacua,
we find that, independent of the value of $\hat c$,  the symmetric
$S^{10}$ de Sitter universe has the largest tunneling probability, with 
${\F} \simeq 10^9$. If the tunneling is via a $S^4 \times M$  
with a pair of $D$9-brane-$\bar D$9-brane, only the 3 spatial 
dimensions in $S^4$ will grow exponentially (inflate), while the 
other dimensions are cosmologically stabilized, i.e., time-independent
(though the amount of inflation is negligible due to the 
fast tachyon-rolling). 
Introducing the dilaton does not change these qualitative features.

\begin{figure}
\begin{center}
\epsfig{file=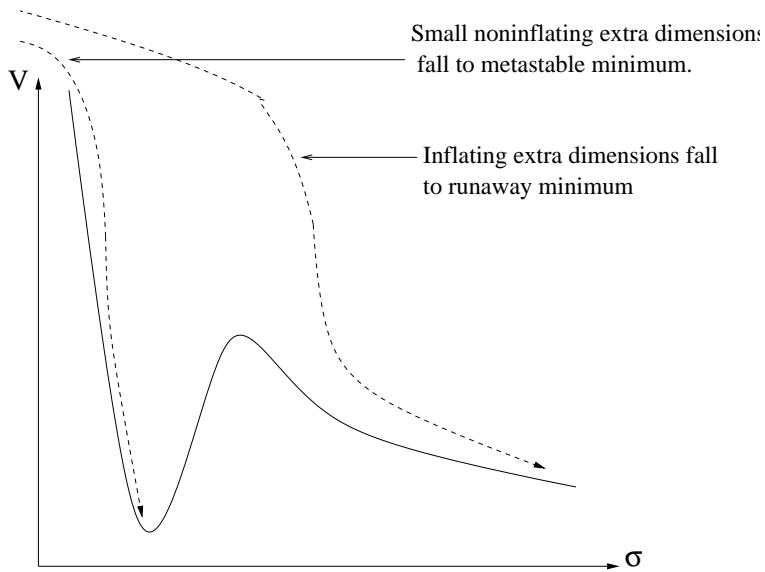, width=10cm}
\vspace{0.1in}
\caption{The potential $V(\sigma)$ as a function of the size 
$\sigma$ of the compactified dimensions. If the size of the
extra dimensions grows substantially in the early universe, the 
universe will end up with 10 uncompactified dimensions.
So cosmological stabilization may be needed in addition to 
dynamical moduli stabilization.
}
\label{fig4}
\end{center}
\end{figure}

Even if we assume that the moduli are all stabilized in today's universe,
there is no guarantee that they should be stabilized during 
the inflationary epoch. (See Figure \ref{fig4}.) 
In KKLT, the K{\"a}hler moduli
are only metastabilized; that is, if they start out with relatively 
large values, they will continue to grow and the universe decompactifies
to a higher dimensional universe. (In the simple case of a single 
K{\"a}hler modulus, the universe decompactifies to 10 dimensions.) This 
runaway solution is always present in superstring theory. So, if the 
extra dimensions grow during inflation, they may easily grow by a 
big enough factor and decompactification takes place.
The presence of cosmological stabilization of the extra dimensions
ensures that the K{\"a}hler moduli are not growing, so towards the end 
of inflation, they can fall down to the locally stabilized values, 
as indicated in Fig. \ref{fig4}. 
In our scenario, universes with large sizes 
of the extra dimensions are suppressed by the decoherence term.
After tunneling, they remain unchanged during inflation. Their 
stabilization can happen before, during or immediately after inflation.

We then consider the brane inflationary scenario \cite{Dvali:1998pa}
as realized in string theory by Kachru et. al. \cite{Kachru:2003sx}. 
When we apply the wavefunction $\Psi$ to this 
scenario, it actually selects a specific inflationary scenario with 
the fluxes determined. That 
is, after inflation, the universe settles down to a specific vacuum 
state, where all the fluxes (say $(M_1,K_1)$) and the moduli are fixed.
One may consider tunneling directly to such a KKLT vacuum state without
going through inflationary stage. Such a preferred vacuum state 
will have different fluxes (say $(M_0,K_0)$).
We find that tunneling to such a vacuum state without going through 
inflation is suppressed relative to the tunneling
to an inflationary vacuum state which then ends up in the $(M_1,K_1)$
state. In the search of possible vacuum
state for us to live in, this implies that the $(M_0,K_0)$ vacuum 
state is not preferred. Of course, subsequent rolling \cite{Kofman:2004yc}
and tunneling from the  $(M_1,K_1)$ state to another $(M_2,K_2)$ 
state are still possible \cite{Feng:2000if,Bousso:2000xa}.
After a discussion on supercritical string vacua, we conclude with some remarks.

Our argument for the decoherence term is admittedly crude and naive.
However, further refinement of the wavefunction $\Psi$ is likely to 
retain some of these qualitative properties. Examination of more 
possible string vacua will help us to pin down the properties of 
$\Psi$. This approach offers the hope that SOUP will select a 
particular inflationary universe that evolves to a specific vacuum 
and thus the anthropic principle may be avoided.

Many of the details are relegated to the appendices.

\section{Wavefunction of the Universe and Tunneling Probability}

The qualitative picture of quantum tunneling from nothing to 
an inflationary universe is relatively well accepted. However, the 
correct wavefunction of the universe, or equivalently, the formula
for the tunneling probability, has remained controversial. 
Here we give a brief introduction (see Appendix A for 
a sketchy review). We then present our proposal on the improved wavefunction 
and the corresponding tunneling probability. We first discuss the 
issues for a de Sitter instanton in 4 dimensions. We then generalize 
the discussion to 10 dimensions. Finally, we propose a 
phenomenological realization of the decoherence effect to be 
incorporated into the wavefunction.

\subsection{Birth of de Sitter Universes}

Let us first review the creation of an inflationary universe in 
4 dimensions. (The generalizations to other dimensions are 
straightforward and will be discussed later.) 
We start with the action for a closed universe $M$,
\be
\label{action4}
S= \frac{1}{16 \pi G} \int_M d^4x \sqrt{|g|} \left(R - 2 \Lambda \right)
+\frac{1}{8 \pi G} \int_{\partial M} d^3x \sqrt{|h|} K
+\int_M d^4x \sqrt{|g|} L_m (\phi)
\ee
where $\Lambda \ge 0$ is the cosmological constant and $L_m$ is the
Lagrangian for all matter fields labeled as $\phi$s. 
Here $h_{ij}$ is the 3-metric 
on the boundary $\partial M$ and $K$ is the trace of the second 
fundamental form of the boundary. 
The extrinsic curvature $K$ in the York-Gibbons-Hawking surface 
term will be important for our discussion.
Here the universe is assumed to 
be closed, homogeneous and isotropic.
To get the picture of the tunneling, consider Figure 1, 
with metric 
$ds^2= N(\tau)^2d\tau^2 + a(\tau)^2 d {\Omega}_3^2$,
where the scale 
factor $a(t)$ in the de Sitter region with lapse function $N=-i$ 
(where $t=N\tau=-i \tau$) is the solution of the Einstein 
evolution equation :
\be
\label{evolutioneqM}
{\dot a}^2 +1 =H^2 a^2,  
\ee
where $H$ is the Hubble parameter,
$H^2 = \Lambda/3$.
This describes the de Sitter space :
\be
\label{dSinstanton}
a(t) = \frac{1}{H} \cosh (Ht)
\ee
in the absence of matter field excitations.
For positive $t$, this describes the inflationary universe.
The Euclidean version of the action (\ref{action4}) can be obtained 
by replacing $t \rightarrow -i\tau$. This gives the Euclidean 
version of Eq.(\ref{evolutioneqM}) :
\be
-{\dot a}^2 +1 = H^2 a^2 
\ee
which gives the $S^4$ solution:
\be
\label{Eucsol} 
a(\tau)= \frac{1}{H} \cos (H \tau)
\ee
This is the well-known de Sitter instanton, which is a compact space.
It is half a four-sphere and is defined only for $|\tau| \le \pi/2H$.
This instanton is interpreted as the tunneling from nothing 
(i.e., no classical space-time) to an inflationary universe
at $\tau = 0$ with size $a=1/H$ and $\dot a=0$, as shown in Figure 1.
We expect the vacuum energy to come from an inflationary 
potential $V(\phi)$, with its (local) maximum at $\Lambda$.

The curvature of the $S^4$ instanton solution is $R=12H^2$.
For $G_N\Lambda << 1$, semiclassical approximations should be valid 
and we may calculate the tunneling probability following standard 
field theory procedures: $P_{HH} \simeq e^{-S_E}$
where $S_E$ is the Euclidean action of half a $S^4$ instanton. 
However, 
\be
\label{SE}
S_E=-\frac{3 \pi}{2 G_N\Lambda}
\ee
which is negative. (The entropy $S_{entropy} = -S_E$.) 
In fact, the Euclidean action for gravity
is unbounded from below. This requires a closer look at the 
tunneling probability.
  
Following Hawking \cite{Hawking:1980gf}, 
let us start with the wavefunction 
of the universe $\Psi$ in the path integral formalism. 
For spatially closed universes, one may express
$\Psi [h_{ij}]$ as
\be
\Psi[h_{ij}] = \int_{\emptyset}^{h_{ij}} Dg_{\mu \nu} 
e^{-S[g_{\mu \nu}]}
\ee
where $h_{ij}$ is the 3-metric and matter fields are ignored
for the moment to simplify the discussion.
Here $S$ is the classical Euclidean action.
The functional integral is over all 4-geometries with a space-like
boundary on which the induced metric is $h_{ij}$ and which to
the past of that surface there is nothing (see Figure 1).

Consider the tunneling from nothing to half a four-sphere 
($\tau=-\pi/2H$ to $\tau_0=0$ with $a=1/H$) and then 
the universe evolves classically in the inflationary epoch 
($\tau > \tau_0$ to ${\hat a}>1/H$,
as indicated in Figure \ref{fig1}).
The action has the value (see Appendix B):
\ba
\label{SDpS}
S(\hat a)=S_R +i S_I= -\frac{3 \pi}{2 \Lambda}
\left( 1 \mp i((H^2{\hat a}^2-1)^{\frac{3}{2}} \right) \ .
\ea
We may view the real part $S_R$ as the Euclidean action due to 
tunneling while the imaginary part $S_I$ as the phase change due 
to the classical evolution of the inflationary universe. 

It is more convenient to let $\tau$ be a parameter or coordinate
($0 \le \tau \le 1$),
so the transition from Euclidean to Lorentzian time is facilitated 
by the transition of the lapse function from real ($N=1$) to 
pure imaginary ($N=-i$) values. 
Once we are ready to entertain complex $N$, we should include 
the tunneling from nothing directly to an universe with size $\hat a$,
where $H {\hat a} >1$.
In this case, the path integral is 
\ba
\Psi(a)=\int_C dw dz \mu(z,a,w) \exp(-S(z,a,w))
\ea
where $w$ symbolically labels the different paths in the 
complex time $T$-plane, and $z$ parametrizes
the instanton (see Appendix B for details). It turns out
that $S(z,a,w)=S(z,a)$. Using the steepest-descent method, where the
steepest-descent paths are the paths with $S_I=$constant,
we find again the above action (\ref{SDpS}), where 
the saddle points correspond to  $H {\hat a} >1$.

Next we consider the ten dimensional case, which corresponds 
closer to the situation in string theory. The analysis for $S^{10}$
is very similar to that of the above $S^4$ case. Other cases such
as the $S^4 \times S^6$ (or more generally 
$S^{1+n_1} \times S^{n_2}$) instanton does not differ much from the 
$S^4$ case; although the $S^4$ component becomes the de Sitter space,
the size of the $S^6$ is fixed, with its value dictated by the same
$\Lambda$ (see next section and Appendix E for more discussions).
The case of $S^4 \times M$ where the compactification of $M$ is 
dynamically stabilized is also quite similar to the above case.
The case that is interestingly new (from our perspective)
is when the size of $M$ is not fixed.
One may consider either the case where the size of $M$ also varies
during inflation (e.g., Kasner-like solutions), or the case where the 
size of M is static during inflation. 
In either case, the result we are looking for does not 
differ much. Since the latter case is more novel,
we shall focus on the $S^4 \times M$ case 
where $M$ is cosmologically but not dynamically stabilized.
To be specific, let us consider the case of 
$S^{1+n_1} \times T^1 S^{n_2} \times T^m$ where $T^1$ is fibered 
over $S^{n_2}$. The Einstein equation is easy to solve. 
See details in Appendix E. To match onto the earlier discussion, let
us consider $S^4 \times T^1 S^2 \times T^3$.
We find that, after tunneling, $S^4$ turns into de Sitter space 
with exponentially growing $a(t)$, while
both the $S^2$ radius and the torus radii are constant (see Figure 4). 
That is, they are cosmologically stabilized during the inflationary 
epoch.
Although the $S^2$ radius is fixed in terms of $\Lambda$, the torus 
radii are undetermined. If they start out too big, after inflation 
they will enter the regime shown in Figure 4 where they will simply 
grow and become decompactified.

Following a similar analysis for the above $S^4$ case, we find
that, for the 10-dimensional vacuum and $H {\hat a} > 1$,
after summing over all degenerate paths:
\ba
\label{SDpSV6}
S(\hat a)=S_E +  i S_I= -\frac{90 \pi V_6}{42 G_{10}\Lambda}
\left( 1 \mp i((H{\hat a})^2-1)^{\frac{3}{2}} \right) 
\ea
where $G_{10}$ is the ten-dimensional gravitational constant,
$V_6$ is the compactification volume of $T^1 S^2 \times T^3$,
$H^2 = 7\Lambda /30$ and $\hat a$ is the cosmic scale factor
of the de Sitter space immediately after tunneling.
We see that the imaginary part $ S_I$ of $S(\hat a)$  
is generically large if (i) $\Lambda$ is small, (ii)
$\hat a$ is large, and/or (iii) $V_6$ is large.
Generalization to other cases is straightforward.
For a D-dimensional inflationary universe with $n$ dimensional 
compactified volume $V_n$, one has, for ${\hat a} \gtrsim 1/H$,
\be
S({\hat a}) \propto -\frac{{\hat a}^{D/2} V_n \Lambda}{G_{D+n}}
\left( 1 \mp i((H{\hat a})^2-1)^{\frac{3}{2}} \right)
\ee

\subsection{Lifting the Feynman Path Degeneracy}

Following the decoherence criterion, tunneling in
each of the above 3 situations should be suppressed when the 
the effect of matter fields is included.
Note that the above formula seems to suggest that, for 
$H{\hat a}=1$, $S_I=0$ even for large $V_6$ and 
$1/\Lambda$. It is our assertion that 
the lifting of the path degeneracy takes place in a way 
such that generic $S_I$ is small only if each of the
factor is not big. This may be a very strong assumption.

\begin{figure}
\begin{center}
\epsfig{file=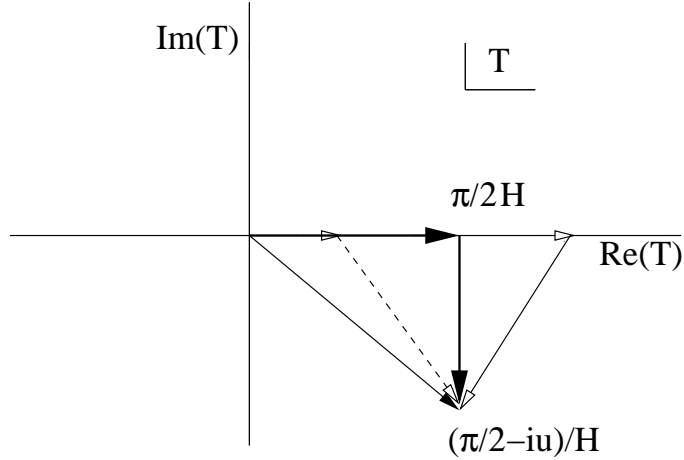, width=9cm}
\vspace{0.1in}
\caption{The 4 paths for $T=0$ to $T= (\pi/2 - i u)/H$ discussed in 
the text, from left to right :
$w=0$, $0<w<\pi/2H$, $w=\pi/2H$ and $w>\pi/2H$. Actually, there are 
infinitely many paths reachable simply by analytic continuation. 
Here $\cosh u = {\hat a} H$.
}
\label{paths4}
\end{center}
\end{figure}

Introducing $T=N\tau$, the integral over $T$ in the action
involves an analytic function
in which the contour in the complex $T$-plane can take a variety
of paths, a subset of which is shown in Figure \ref{paths4}.
It is convenient to shift the $\tau$ coordinate 
$\tau \rightarrow \tau + \pi/2H$, so $\tau$ is always positive.
For fixed $z$, let us consider 4 paths (among infinitely many) 
shown in Figure \ref{paths4} to
illustrate some of the points we would like to make.

\begin{itemize}

\item $w=\pi/2H$. This integration of $T$ along the 
real axis (i.e., Euclidean time : $\tau$ from $0$ to $1$ with 
$N=\pi/2H$) 
corresponds to tunneling to half a four-sphere, and then the
de Sitter universe evolves classically as $T$ goes 
in the imaginary direction (i.e., Lorentzian time) from 
$\pi/2H$ to $(\pi/2 \mp i u)/H$, where $\cosh u = {\hat a} H$.  
This is the only path included in the Hartle-Hawking wavefunction.

\item $w=0$. The integration is along the $T=N\tau$ path for 
$$N = \left[ \frac{\pi}{2} - i u \right]/H$$ 
with $\tau$ going from $0$ to $1$.

\item $0<w<\pi/2H$. This corresponds to tunneling to less than a 
half four-sphere (with $H a <1$), and then continue 
tunneling to $\hat a$.

\item $w>\pi/2H$. This corresponds to tunneling to larger than a
half instanton (with $H a <1$), and then continue
tunneling back to $\hat a$.

\end{itemize}  
These 4 paths (and many others related by analytic continuation
in the complex $T$-plane) all have the same action given by
$S(z,a,w)=S(z,a)$ (\ref{complexS}). The reason they have the 
same action is because of the gauge symmetry coming from
diffeomorphism invariance \cite{Halliwell:1989dy}. 
Ignoring the measure and summing 
over paths via the steepest descent approximation discussed earlier 
and in Appendix B then yields the result (\ref{SDpS}) or 
(\ref{SDpSV6}) as given above.
Let us make a few observations here :

\begin{enumerate}

\item We are interested in an inflationary universe with 
growing $a$. Even if we tunnel to an inflationary universe with
$Ha<1$, $a$ will rapidly grow to $\hat a$ such that 
$H\hat a > 1$. Summing over all paths that end at $\hat a$,
the real part of $\Psi$ goes like
$-3 \pi/2 G\Lambda$ in (\ref{SDpS}) or $\sim V_6/G_{10}\Lambda$
in (\ref{SDpSV6}),
corresponding to tunneling to exactly half an instanton.
This seems to uniquely fix the tunneling probability.
This answer is same as that adopted by Hartle and 
Hawking \cite{Hartle:1983ai} 
in the 4-dimensional case, though the argument is very different, 
since they do not consider paths involving complex metric.

\item So far, the measure and the matter fields have been ignored. 
\ba
S_{eff}(z,a,w) = S(z,a) + S_m(z,a,\phi) - \ln \mu (z,a,w)
\ea
Besides the the measure, which receives contributions from gauge 
fixing as well as other effects, $S_m$ 
includes light matter (both bosonic and fermionic) fields as well 
as excitational modes arising from perturbing the metric. 
These modes/fields are collectively represented by  $\phi$.  
Intuitively, we expect the inclusion of these fields 
to lift the above path degeneracy. This will lead to 
an effective action $S_{eff}(z,a,w)$ that is path-dependent.
Its dependency on $w$ will lead to important effects when the 
imaginary part $S_I$ of the action is large. A particularly 
interesting field to include is the inflaton field in a slow-roll 
inflationary scenario. In this case,
$\Lambda$ is actually the effective inflaton potential $V(\phi)$ and 
the de Sitter description is only approximate. Following the $w=\pi/2H$
path, scale-invariant density perturbation is generated during the
classical evolution of the universe. For other paths, say the $w=0$ path, 
presumably no $\phi$ fluctuation is produced during tunneling.
In any case, whatever density perturbation generated during tunneling is 
expected to be very different from that generated during inflation. This 
illustrates the lifting of the path degeneracy.

\item Suppose $Im~ S({\hat a},w)$ is very 
large; a small lifting of the path
degeneracy due to $S_m$ (and $\mu (z,a,w)$) will introduce a 
perturbative change in $ S(z,{\hat a},w)$, which can still be 
much larger than $\pi$. 
This destructive interference among the paths is akin to decoherence, 
and any quantum effect will be suppressed, leaving behind only 
classical behavior. Since classically there is no tunneling, this 
will suppress the tunneling probability.
Let us assume that decoherence takes place when
$Im~ S(\hat a)$ is very large. Generically this happens in the 
following 3 situations :

\begin{itemize}

\item Very small $\Lambda$. 
A phase variation due to $S_m$ will suppress the
quantum tunneling. That is, in a model where $\Lambda$ is dynamical,
the quantum tunneling to a de Sitter universe
with a very small cosmological constant is suppressed.
With today's dark energy, $(G\Lambda)^{-1} \sim 10^{120}$. 
So a very small lifting of the path degeneracy 
that causes a tiny percentage change in the value on $S$
will be sufficient to suppress the tunneling to a 
de Sitter universe with a very small cosmological constant.

\item $H {\hat a} >>1$. The lifting of path degeneracy  will suppress 
the quantum paths that tunnel directly or close to a large $\hat a$. 
This suggests that tunneling from nothing to a de Sitter 
universe with $a$ not too large (say, close to but a little larger 
than $H^{-1}$) is most likely. The universe then evolves 
classically (i.e., inflates) from such a value of $a$ to
$\hat a$ and beyond. This allows us to preserve the 
almost scale-invariant density perturbation generated 
during inflation.

\item Large $V_6$. Again, a phase variation due to $S_m$ will 
suppress its quantum tunneling. That is, in a model where 
size of the extra dimension is dynamical,
the quantum tunneling to a de Sitter universe
with a very large compactification volume (or in the uncompactified 
limit) is suppressed.

\end{itemize}

\item Since the real part of the action $S_R$ suppresses tunneling 
to a vacuum with a very large $\Lambda$, we expect the desirable
value of $\Lambda$ to take some intermediate value below the Planck scale.
This then 
justifies the semi-classical approximation we have been relying on.

\end{enumerate}

\subsection{An Improved Wavefunction}

In usual quantum mechanics, decoherence occurs when a 
macroscopic object interacts with its environment, e.g., a dust 
particle in the bath of the cosmic microwave background radiation 
in outer space. However, for the universe,
there is no outside observer to play the role of the
environment. Instead, the classical curved metric plays the role of
the macroscopic object, while the perturbative modes of gravity as 
well as the matter fields interacting with classical 
gravity play the role of the environment, as explained by 
Kiefer and others \cite{Kiefer:1987ft,Halliwell:1989vw}. 
Since this point is rather 
important in SOUP, we give a brief review in Appendix C.

The $S_E$ term in $\Psi$ suppresses the tunneling to a large
$\Lambda$ universe. For small $\Lambda$, the universe is large,
i.e., macroscopic, so we expect decoherence to suppress its 
tunneling (since there is no tunneling in classical physics).
Phenomenologically, we expect an inflationary universe with a 
moderate value of $\Lambda$ (which is much larger than today's 
dark energy and smaller than the Planck scale) that evolves 
more or less classically before it ends with the hot big bang 
epoch. In de Sitter space or an inflationary universe,
the cosmic scale size $\hat a$ plays the role of the 
configuration variable. Here we would like to find the leading 
term in $\cal D$ due to the environment.

A mildly path-dependent measure $\mu(z,a,w)$ and $S_m(\phi)$
may allow us to realize this scenario via decoherence.
For what values of $\Lambda$ that this decoherence effect will 
begin to suppress the tunneling is a model-dependent question.
The standard way to control a path integral that involves 
phases is to go to imaginary time. Such a calculation is beyond 
the scope of this paper. Instead, let us take a phenomenological 
approach here to find the functional form of the leading 
contribution to $\cal D$. 
Let us consider two approaches: $\Sp$-brane and decoherence. 
They yield the same functional form for the leading term in $\cal D$.

\begin{itemize}

\item

Although all paths in the Feynman path integral approach 
involve continuous metric, paths with discontinuous derivatives of 
the metric are allowed; that is, the extrinsic curvature $K$ can 
have jumps so that its integral in the action can be finite.
The discontinuity of $K$ corresponds to a time-localized defect. 
In string theory, such ``defects'' are known as space-like branes or
S-branes \cite{Gutperle:2002ai}. 
Since the defect here is located at complex time,
we shall call it $\Sp$-brane. More generally, a $\Sp$-brane has some 
structure, which is reflected in a rapidly changing (but smooth) 
$K$ in some time region. That is, at the location of such a $\Sp$-brane,
the boundary term can be expressed as 
\ba
\label{S'action}
\frac{1}{16 \pi G} \int_{\partial M} d^9x \sqrt{|h|} {K}
 = +T_{\Sp}\int d^9x \sqrt{|h|} \simeq T_{\Sp} V_9
\ea
where $T_{\Sp}$ indicates the $\Sp$-brane tension and $V_9$ 
the 9-spatial volume. For the $w= \pi/2H$ case, $V_9$ is simply 
the ``boundary area'' of the half-ten-sphere. 
One anticipates that the tension of such a $\Sp$-brane to be 
around the string scale (up to factors of $2 \pi$). 
So we are led to the following ansatz:
\ba
\Psi &\simeq \exp {\F} = \exp \left( -S_E - {\cal D} \right)\\\nonumber
{\cal D} & \simeq c V + ...  = c V_9/l_s^{9} +...
\ea
where $c$ is a dimensionless parameter, 
$V$ is the dimensionless spatial volume,
$l_s=2 \pi \sqrt{\alpha^{\prime}}=2\pi/M_s$,
and $V_9$ is the ``area of the boundary'' at the transition from
Euclidean to Lorentzian space (see Figures 1 and 2).
That is, $cV$ is the leading term in the decoherence term $\cal D$.
Here, $c$ may have a mild dependence on the matter fields, both 
open and closed string modes, as well as on the excitational modes 
emerging from perturbing the classical metric. 
The determination of the value of $c$ will be a challenge.
Sometimes it is convenient to introduce ${\hat c}{\hat V}$ 
where $\hat V$ is the ``spatial volume'' of the de Sitter (or any other)
instanton (as measured in string scale $M_s^2= 1/\alpha^{\prime}$);
that is, ${\hat V} = M_s^9 V_9$ for critical string vacua.
So $c=(2\pi)^9 {\hat c}$.

A constant term in $\cal{D}$ may be absorbed into the normalization of
$\Psi$. The $cV$ term suppresses universes with a small
$\Lambda$ and/or a large inflationary size.
To suppress universes with very large (or uncompactified)
extra dimensions, we need higher order terms. Possible terms like
$V_{10}^2$, $V_9 V_{10}$ etc. can easily do the job, but
terms like $S_E^2$, $V_9^2$ etc are not useful. 
Fortunately, a realistic string vacuum will have the compactification
of its extra dimensions dynamically stabilized. In those realistic 
situations, we may assume the higher order terms in ${\cal{D}}$ to 
be negligible. So, with some luck, a phenomenological analysis 
keeping only the lowest order term in Eq.(\ref{master}) may be 
sufficient.

\item 
Now we turn to the quantum decoherence approach 
\cite{Zurek:1982ii,Kiefer:1987ft,Zeh:1988ws}, 
which gives the same leading 
term in ${\cal{D}}$. In quantum cosmology the total wavefunction in a 
Friedmann solution with cosmic scale $a$ is
\be
\label{fullwf}
\Psi (a, {x_{n}})=  \psi_{o}(a)\prod_{n>0}\chi_{n}(a,x_{n})
\ee
where $\psi_{o}(a)$ is (up to a normalization factor) 
the Hartle-Hawking wavefunction, and
$\chi_{n}(a,x_{n})$, with amplitudes $x_{n}$, (an orthonormal set 
at a given value of $a$) are the contributions 
from all other fields. 
Here the metric $a$ is the configuration variable describing the 
semi-classical evolution of the universe while the higher multipoles 
of gravity and other matter fields interacting with gravity
provides the environment.
These fields interact with the gravitational contribution 
encoded in $\psi_{o}(a)$ and the effect of this interaction can 
be seen by considering the reduced density matrix, where the 
environment is traced over:
\be
\rho_S(a ; a^{\prime}) = tr_{x_{n}} \left(|\Psi><\Psi| \right) 
= \psi_{o}(a)\psi_{o}^{*}(a^{\prime}) \prod_{n>0} \int_{-\infty}^{\infty}
dx_{n}\chi_{n}(a, x_{n})\chi_{n}^{*}(a^{\prime}, x_{n})
\ee
For tunneling from nothing ($a^{\prime}=0$) to $a$, we see that 
$\Psi (a) \simeq \rho_S (a ; a^{\prime}=0)$.
In the absence of the environmental degrees of freedom, 
$\Psi$ reduces to the Hartle-Hawking wavefunction.
Let us first consider the finite $a^{\prime}$ case.
In the presence of the environment, the reduced density matrix 
takes the form 
\be
\rho_S (a ; a^{\prime}) \simeq \psi_{o}(a)\psi_{o}^{*}(a^{\prime}) 
\prod_{n>0} \exp \left( - f(a,a^{\prime})(a-a^{\prime})^2 \right)
\ee
Contributions to $f(a,a^{\prime})$ from
the tensor modes of the metric \cite{Kiefer:1987ft}, 
massless minimally coupled inhomogeneous scalar 
fields \cite{Halliwell:1989vw},
fermionic modes \cite{Kiefer:1989ud},
Kaluza-Klein modes \cite{Mellor:1991mx}
and conformally coupled scalar fields \cite{Kiefer:1992cn}
have been studied when both non-zero $a$ and $a^{\prime}$ are in the 
classically allowed region. These works justify the validity of the
classical time evolution of the universe. Here we use it to 
obtain the form of the leading decoherence term.

All these modes yield very similar results.
To be specific, let us focus on the massless scalar field
in the $S^4 \times M$ case, extending the result to the 
under-the-barrier region.
Each of its modes contributes $(a+a^{\prime})^2/4a^2a^{{\prime} 2}$ 
to $f(a,a^{\prime})$ \cite{Halliwell:1989vw}, so we have
\be
\label{den}
\rho_S(a ; a^{\prime}) \propto 
\prod_{n>0}
\exp \left(- \frac{(a+a^{\prime})^2(a-a^{\prime})^2}
{4a^2a^{{\prime} 2}} \right) \sim 
\exp \left(- N\frac{(a+a^{\prime})^2(a-a^{\prime})^2}
{4a^2a^{{\prime} 2}} \right)
\ee
where $N$ is the total number of modes included in the environment.
Taking the limit of infinitely many modes would reduce the Gaussian
distribution to a $\delta$-function, corresponding to an exact 
diagonalization of the density matrix. 
This means the scale factor $a$ has been perfectly measured, 
which in turn implies that its momentum conjugate $ -{\dot a} a$ 
(or simply ${\dot a}$) must 
have an infinite spread, which is highly non-classical (a squeezed
state). For $N=0$, i.e., no modes contribute, $a$ would have
an infinite spread, a rather unrealistic, highly non-classical 
situation.
So, on physical grounds, one expects a cut-off on the modes so that
the expoenent in $\rho_S$ (\ref{den}) is finite. 
Clearly we need a regularization scheme. 
Fortunately, string theory provides a natural cut-off.
Modes with wavelengths larger than the horizon size are not 
observable.
Consider the above $S^4 \times M$
case. The higher multipoles can be found from an expansion of the 
full 3-dimensional spherical harmonics (the boundary $S^3$ of $S^4$)
and normal modes in a toroidal $M$. 
So $n \rightarrow (n,l,m, n_1, n_2, ...,n_6)$
where $(n,l,m)$ is for $S^3$ and the remaining $n_i$ are for $M$. 
It is reasonable to only trace out modes with long 
wavelength $\lambda$, (say  $> {\bar a}/n$ in boundary $S^3$,
with ${\bar a}$ being some average of $a$ and $a^{\prime}$, 
and $ \ge  L/n_i$, where $V_6 \simeq L^6$), 
we see that, $N \propto V_6 {\bar a}^3 \simeq V_9$.
Consider a field with wavenumber
$k \equiv 2 \pi/\lambda = 2 \pi n/L$ on a torus with volume $L$. 
Since $k$ is cut off at $k_{max} \simeq 2 \pi/l_s$, the number of 
modes that contributes is $N \simeq L/l_s$. Generalizing this to 
a higher compactified $d$-dimensional volume $V_d$, we have 
$N \simeq V_d/l_s^d$. For $4$-dimensional de Sitter space in 
$D$-dimensional spacetime, we have  
\be
{\cal D}  = c N
 = c (V_{3}/l_s^3)(V_{D-4}/l_s^{D-4})
 =  c V_{D-1}/l_s^{D-1}
\ee

Now consider the case we are interested in, that is, the limit 
$a^{\prime} \rightarrow 0$ in 10 dimensions. 
In this limit, all modes are 
essentially frozen. The validity of a semi-classical description
of the universe requires finite spreads in both $a$ and $\dot a$. 
To avoid either an infinitely spread in $a$
or an infinitely spread in $\dot a$, we must have
$N \sim V_6 {\bar a}^3 \sim V_6 a a^{{\prime} 2}$.
So for  $a^{\prime} \rightarrow 0$ and 
$a \simeq H^{-1} \simeq \Lambda^{-1/2}$, we have
\be
{\cal D} \propto \frac{(V_6 a a^{{\prime}2}) (a^2-a^{{\prime} 2})^2}
{a^2 a^{{\prime} 2}} 
\rightarrow \frac{V_6}{\Lambda^{3/2}} \simeq V_9
\ee

By considering a few universes other than $S^4 \times M$, 
we see that the decoherence argument agrees with 
the $\Sp$-brane argument, that is
\be
\Psi \simeq \psi_{o}^{*}(0) \exp \left(-S_E -  c V_9/l_s^9 \right) 
\ee
where $\psi_{o}^{*}(0)$ provides the normalization.

Note that the actual determination of $\cal D$ requires knowing 
the full wavefunction $\Psi(a=0)$ (\ref{fullwf}): this is the 
wavefunction when there is no classical spacetime. 
Since classical spacetime is fundamental in general relativity, 
$\Psi(a=0)$ may not be well-defined in quantum gravity. It will be 
interesting to see if string theory can fix $\Psi(a=0)$, or it has 
to be postulated. The parameter $c$
actually depends on the particular vacuum in string theory.
It is presumably a good approximation to take $c$ to
be constant for a subclass of vacua.
\end{itemize}
The inclusion of $\cal D$ should suppress instantons with more 
complicated geometries, such as those considered in 
Ref\cite{Fischler:1990se}. 
Higher order terms involving higher powers of the Riemann tensor and field 
strengths are present in low-energy effective theory in string theory, so
they should be included in $S_E$ \cite{Hawking:2002af,Lu:2004fe}.
In the rest of this paper, we shall ignore such terms and the higher 
order terms in ${\cal{D}}$ as well as the variation of $c$.
This is certainly sufficient for illustrative purposes.

\section{Toy Models : Ten Dimensional Gravitational Instantons}

To begin, we study a few instanton solutions in ten dimensions. 
We will be interested in $S^{10}$ as well as $S^4 \times M$ solutions 
in which four ($1$ time and 
$3$ space after Wick rotation) out of the ten dimensions inflate 
while $M$ of the remaining six spatial dimensions have static behavior
during an early inflationary phase. 
These different solutions will have different topologies for 
the extra six dimensions. 

\subsection{$S^{10}$}

This instanton solution is a trivial generalization of the $S^{4}$ 
instanton. The Euclidean metric ansatz for $S^{1+n}$ is:
\baray
ds^{2} =  dt^{2} + a(t)^{2}\left(  \frac{dr^{2}}{1-r^{2}} +  r^{2} d\Omega^{2}_{(n-1)} \right)
\earay
The Euclidean Einstein equations for $S^{1+n}$, for any $n$, are:
\baray
-\frac{1}{2}n(n-1) \left( \frac{\dot{a}^{2}}{a^{2}} - \frac{1}{a^{2}}\right) = \Lambda \\ \nonumber
-(n-1)\frac{\ddot{a}}{a}- \frac{1}{2}(n-1)(n-2)  \left( \frac{\dot{a}^{2}}{a^{2}} - \frac{1}{a^{2}}\right) = \Lambda
\earay
The bounce solution is given by $a(t) = H^{-1}\cos(Ht)$ with
$H^{2}= {2\Lambda}/{n(n-1)}$ and $R=n(n+1)H^2=2(n+1)\Lambda/(n-1)$.
For the case of $S^{10}$, this becomes $H^{2}= \Lambda/36$. 
By Wick rotating the time axis one gets a closed $9$ spatial dimensional 
inflating universe with the scale factor given by $a(t) = H^{-1}\cosh(Ht)$.

Now, in string theory, $\Lambda$ can come from the presence of 
extra branes. Starting with a supersymmetric vacuum, we can add N
pairs of $D9-{\bar D}9$ branes, where the vacuum energy $\rho_{vacuum}$ 
is just 2$N$ times the $D9$-brane tension,
\baray
\label{vac}
\Lambda =N \Lambda_1=8 \pi G_{10} \rho_{vacuum} = 16 \pi G_{10} N 
\frac{M_{s}^{10}}{(2\pi)^{9}g_{s}}
\earay
where $g_{s}$ and $M_{s}=1\sqrt {/\alpha^{\prime}}$ are the string 
coupling constant and the string mass scale, respectively.
For numerical calculations, we shall take $g_{s}=1$.
The ten dimensional Newton's constant is given by
\be
\label{newton}
 8\pi G_{10} = \frac{g_{s}^{2}(2\pi)^{7}}{2 M_{s}^{8}}
\ee
The entropy of a de Sitter space equals the Euclidean action, 
$S_{entropy}=-S_E$. In the presence of matter fields, matter
excitation modes can be present. However, it is known that
the pure de Sitter space saturates the entropy bound 
\cite{Bousso:2002fq}, so we expect the pure de Sitter instanton
to have the largest tunneling probability. 

In string theory, the dilaton plays a crucial role in the gravity 
sector. A summary of the analysis including the dilaton
can be found in Appendix D. The inclusion of the dilaton field does 
not change the qualitative features we are interested in.

\subsection{$S^{4} \times S^{6}$, $S^n \times M$ etc.}
 
We start with the 10-dimensional Euclidean metric ansatz
for $S^{1+n_{1}} \times S^{n_{2}}$ :
\baray
\label{Sn1Sn2}
ds^{2}=  d\tau^{2} + a(\tau)^{2}\left( \frac{dr^{2}}{1-r^{2}} +
r^{2}d\Omega^{2}_{(n_1-1)}\right) + b(\tau)^{2} \left( \frac{d\rho^{2}}
{1-\rho^{2}} + \rho^{2}d\Omega^{2}_{(n_2-1)}\right)
\earay
The solution of the corresponding Einstein equation requires 
$b$ to be constant (see Appendix E) with
\ba
b^2~&=&~ \frac{(n_{2}-1)(n_{1}+n_{2}-1)}{2\Lambda} \\ \nonumber
H^{2} ~&=&~ \frac{2\Lambda}{n_{1}(n_{1}+n_{2}-1)}
\ea
For $n_1=3$, this yields a 4-dimensional de Sitter space and 
a static $S^6$. That is, if we consider time to lie in $S^{4}$ after Wick 
rotation, then it describes a ten dimensional universe with inflation 
occuring in the $3+1$ dimensions corresponding to $S^{4}$ while the
remaining $6$ spatial dimensions in $S^{6}$ remain static. If we 
include a small time varying component to $b(\tau)$, it will be 
rapidly inflated away.

It is easy to generalize the above analysis to other cases. In general,
for $S^{n_{1}} \times S^{n_{2}} \times S^{n_{3}} \times ... $, when 
rotated to Lorentzian time, only the spatial directions in $S^{n_{i}}$
which contains the time direction will inflate, while all the 
rest of the spatial directions remain static. 

We tabulate the probabilities thus calculated for the different geometries.
(The list is incomplete and is representative only.)

\vspace{0.3cm}

\begin{tabular}{|c|c|c|c|c|} \hline
Geometry       &   $-S_E \times 10^{-9}$    &  Volume 
$\times 10^{-15}$ &   $\frac{N_m}{{\hat c}^{2}} \times 10^{-12}$ & 
${\F} {\hat c}^{8}\times 10^{40}$   \\ \hline\hline
$S^{10}$   &   $3.94170$   &  $3.92236$  & $1.25626$ & $1.75627$\\
$S^{8} \times S^{2}$ &   $2.88416$   &  $2.89347$ & $1.27620$& $1.20694$ \\
$S^{7} \times S^{3}$ &   $3.07644$  &  $3.11670$ &  $1.29131$ & $1.23300$\\
$S^{6} \times S^{4}$ &   $3.17258$  &   $3.20829$ &  $1.31267$& $1.17892$\\
$S^{5} \times S^{5}$ &  $3.17258$  &   $3.26494$ & $1.34511$ & $1.07489$\\
$S^{5} \times S^{3} \times S^{2}$ &  $2.24004$   &  $2.30866$ & $1.34511$ &
$0.760062$\\
$S^{4} \times S^{6}$ &  $3.17258$ & $3.31351$ & $1.40018$ & $0.910688$  \\
$S^{4} \times S^{3} \times S^{3}$ &  $2.40348$  &  $2.49832$ & $1.40018$
& $0.686643$ \\
$S^{4} \times S^{2} \times S^{2} \times S^{2}$ &  $1.50938$  & $1.59048$& $1.40018$& $1.07489$ \\
$S^{3} \times S^{2} \times S^{2} \times S^{3}$ &  $1.58630$  &  
 $1.73150$ & $1.51325$& $0.335524$\\
$S^{2} \times S^{6} \times S^{2}$ &  $2.11506$  &  $2.55074$ & $1.86691$
& $0.192098$\\
$S^{2} \times S^{4} \times S^{4}$ &  $2.26888$  &  $2.75480$ & $1.86691$& 
$0.207466$ \\
$S^{2} \times S^{2}\times S^{2}\times S^{2}\times S^{2}$ &  $1.00946$  &  
$1.22435$ & $1.86691$& $0.092207$ \\ \hline
\end{tabular}

\vspace{1.0cm}

\noindent Table 1. The time coordinate is part of the first factor $S^{n}$. 
The Euclidean action $S_E$ is for a single pair of $D9-{\bar D}9$ branes.
The 9-dimensional volume of the boundary is measured in units of string 
scale, that is Volume$=M_s^9V_9$. The value of $N_m$ is for 
maximizing $\F$, the value of which is given in the last column.
For small $c \equiv (2\pi)^9 {\hat c}$, integer $N_m=1$ maximizes $\F$, 
which has its 
value very close to $-S_E$, not that given in the last column.  

\vspace{1.0cm}

For very small $\hat c$, $N_m=1$ is the optimal choice (since $N_m$ is 
an integer); that is,
one brane pair maximizes $\F$. In this case, we have ${\F} \simeq -S_E
\sim 10^9$, independent of the value of $\hat c$.
Let us make some comments here :

\begin{enumerate}

\item In string theory, there are tunnelings to many possible 
expanding universes. So the relative tunneling probabilities 
become an important issue to pick out the most likely universes. 

\item 
There are at least 2 possible ways to stabilize the moduli in 
superstring theory: 
dynamical versus cosmological. Cosmological stabilization 
has been explored in the presence of string winding modes 
\cite{Brandenberger:1989aj} and more
recently in the presence of branes 
\cite{Alexander:2000xv,Watson:2003gf,Easther:2002mi,Easther:2002qk}. 
Dynamical stabilization has made 
substantial progress recently \cite{Kachru:2003aw,Saltman:2004sn}. 
In this scenario, the universe 
is sitting at a metastable vacuum, with a tiny positive cosmological 
constant. As pointed out earlier, there is a true (supersymmetric)
vacuum with zero cosmological constant where all the extra dimensions 
are decompactified.
Even if we believe in dynamical stabilization, the universe must 
avoid an early phase where the extra dimensions go through a rapid 
growth (as illustrated in Figure 4). So cosmological stabilization 
as presented above may be necessary even in the presence of eventual
dynamical stabilization. 

\item The simplest instanton solution is the de Sitter $S^{10}$ instanton. 
Such a solution represents the tunneling of the universe 
to a space-filling $D9-\bar{D9}$ system in which all the $9$ spatial 
dimensions inflate as the brane and the antibrane decay down the tachyon 
potential. The inflation offered by the $D9-\bar{D9}$ system is, however, 
very marginal, amounting only to a fraction of an e-fold. As this system 
decays into  pairs of $D7-\bar{D7}$ and radiation, the $9$ spacial 
dimensions 
undergo a radiation dominated expansion and this leads to a significant 
separation between the $D7-\bar{D7}$ pairs. This sizeable separation may
lead to a significant number of e-folds in the ensuing inflation between 
the $D7-\bar{D7}$ system.

\item If we use Linde/Vilenkin's wavefunction (see Appendix A for a brief 
review), $\psi \sim e^{-|S_E|} \sim e^{-10^{9}/N}$, so it seems 
that $(S^2)^5$ among the list in Table 1 is preferred in this case.
However, an inflationary universe with large $N$ (i.e., large cosmological
constant) is also preferred. Of course, for large $N$, the 
semi-classical approximation breaks down and we lose all 
control of the estimate. Furthermore, the inclusion of the decoherence term 
would not help.

\end{enumerate}

Other interesting instanton solutions involve geometries of the form 
$S^{m} \times T^1S^{n} \times T^{k}$ are discussed in Appendix E.

\subsection{An Application of the new Wavefunction}

Consider the function $\F$.
If the cosmological constant is provided by $N$ $D9$ brane-antibrane 
pairs, $\F$ can be written as:
\baray
{\F} = \frac{1}{16 \pi G_{10}}\frac{N \Lambda_1}{2}V_{10} - {\hat c}M_{s}^{9}V_{9}
\earay
Considering geometries of the form $S^{1+n_{1}} \times S^{n_{2}}$:
\baray
{\F} = \frac{N \Lambda_1}{32 \pi G_{10}} V_{1+n_{1}} V_{n_{2}} - 
{\hat c}M_{s}^{9} V_{n_{1}} V_{n_{2}}
\earay
where  $V_{1+n_{1}}$ and $V_{n_{2}}$ are the surface areas of 
$S^{1+n_{1}}$ and $ S^{n_{2}}$, respectively.  
$V_{n_{1}}$ is the area of the equator of $S^{1+n_{1}}$. 
One should note that the surface areas depend on $N$ via the Hubble radius,
$H^{-1}$. To find the value of $N=N_m$ at which $\F$ is a maximum,
we just require that $\partial {\F} / \partial N = 0$. 
This gives the following value for $N$:
\baray
\frac{N_m}{{\hat c} ^{2}}= \frac{81}{64}(2\pi)^{16}\frac{\nu(n_{1})^{2}}{n_{1}
\nu(1+n_{1})^{2}}
\earay
where $\nu(n)$ is the ``surface area'' of a unit $n$-sphere:
\baray
\label{area}
\nu(n) = 2\pi^{(1+n)/2}/\Gamma ((1+n)/2)
\earay
One can plug this value of $N$ back in the expression for $\F$ and 
get the extremized value of $\F$:
\baray
{\F}_{max} = \frac{2^{33}}{3^{18}(2\pi)^{63}}
(\frac{n_{2}-1}{n_{1}})^{n_{2}/2}n_{1}^{9}\frac{\nu(n_{2})
[\nu(1+n_{1})]^{9}}{[\nu(n_{1})]^{8}}\frac{1}{{\hat c}^{8}}
\earay
Though all geometries of type $S^{1+n_{1}} \times S^{n_{2}}$ have the 
same dependence of $\hat c$ (viz. $1/{\hat c}^{8}$), they have different 
coefficients in front of  the ${\hat c}^{-8}$ term.  In fact, the volume 
term (${\hat c}M_{s}^{9}V_{9}$) lifts the degeneracy between 
the $S^{m} \times S^{n}$ and the $S^{n} \times S^{m}$ geometries.

We see in Table 1 that, irrespective of the value of $\hat c$, 
tunneling via
$S^{10}$ has the largest probability. This is not surprising, since 
$S^{10}$ is the most symmetric instanton. $S^{10}$ tunneling implies 
that the universe begins with all its 9 spatial dimensions uncompactified.
Including the dilaton will not change this qualitative feature. 
Fortunately, as we shall see that, in string theory, the tunneling 
to a realistic inflationary scenario with 6 dimensions compactified 
is preferred over $S^{10}$. 

A complete understanding of quantum gravity is lacking. However, as 
long as the quantum corrections are small, one can justify the 
semiclassical treatment that we have employed. The curvature of the 
instanton solutions is given by
$R = 5\Lambda/2 = \left(\frac{5N}{8\pi^{2}}\right)g_{s}M_{s}^{2}$. 
The semiclassical description is reliable as long as 
${R}/{M_{s}^{2}} \simeq {5Ng_s}/{8\pi^{2}}<< 1$,
or $Ng_s << 15$. This implies that, for $g_s \simeq 1$, $N \sim 1$. 
In Table 1, we see that
\ba
N_m \simeq {\hat c}^2 \times 10^{12}
\ea
For the above semiclassical analysis to be meaningful, this condition
requires $\hat c \lesssim 10^{-6}$, or
\ba
 c \equiv (2 \pi)^9 {\hat c} \lesssim 10
\ea
which we shall assume is satisfied. 
For $c \ll 10$, we have $N_m=1$ (since $N$ is quantized
and $N=0$ is ruled out). In this case,
the value of $\F$ is actually very close to $-S_E$, since
the $\cal D$ term is completely negligible. As we shall see, the 
phenomenological analysis below puts $ c \simeq 10^{-3}$.
In any case, the $S^4 \times M$ instanton like those in Table 1
is never preferred.

\section{Tunneling to Inflationary Universes in String Theory}

Now we turn to more realistic vacua in string theory.
First, we consider tunneling to string vacua that 
initiate an inflationary phase. We then consider
tunneling directly to string vacua that mimic our universe today
(i.e., with a very small cosmological constant).
To be specific, we shall study the model due to Giddings etc.
\cite{Giddings:2001yu} and Kachru etc. \cite{Kachru:2003aw,Kachru:2003sx}.
It will be important to study as many stringy vacua as possible.
However, the analysis here should be sufficient to illustrate
our approach.  

Start with a 4-fold Calabi-Yau manifold in F-theory,
or, equivalently, a type IIB orientifold compactified on a
3-fold Calabi-Yau manifold 
with fluxes to stabilize all but the volume
modulus \cite{Giddings:2001yu}. For large volume, supergravity
provides a good description.
The presence of $D7$-branes introduces a non-perturbative
superpotential $W$ that stabilizes the volume modulus in a
supersymmetric AdS vacuum. The introduction of $\bar D$3-branes
in a warped type IIB background
breaks supersymmetry and lifts the AdS vacuum to a metastable
de Sitter (dS) vacuum \cite{Kachru:2003aw} (the KKLT vacuum).
To realize inflation, KKLMMT \cite{Kachru:2003sx} introduces 
a $D$3-$\bar D$3-brane pair, whose vacuum energy drives inflation
\cite{Burgess:2001fx,Alexander:2001ks,Dvali:2001fw,Buchan:2003gx}.
The $\bar D$3-brane is fixed with the other $\bar D$3-branes
in the Klebanov-Strassler deformed conifold \cite{Klebanov:2000hb}
and the inflaton $\phi$ is the position (relative to the 
$\bar D$3-branes) of the $D$3-brane.
This yields a potential for the mobile $D$3-brane in this
$D$3-${\bar D}$3-brane inflationary scenario
\ba
\label{fullV}
V(\rho,\phi) &=& V_F({\cal K}(r),W(\rho,\phi)) + \frac{D}{r^2} +
V_{D{\bar D}}(\phi)
\ea
where $r$ is the physical size of the compactified volume and
$\rho$ the corresponding bulk modulus.
$V_F({\cal K}(r),W(\rho,\phi))$ is the F-term potential, where the
superpotential $W$ is expected to stabilize the volume modulus
in an AdS supersymmetric vacuum, while the $\bar D$3-branes
(the $Dr^{-2}$ term) in the warped geometry breaks
supersymmetry to lift the AdS vacuum to a de Sitter vacuum (a
metastable vacuum with a very small cosmological constant and a
lifetime larger than the age of the universe). Note that this 
non-perturbative interaction term leaves $\phi$ to be essentially 
massless, 
that is, there exists a shift symmetry \cite{Firouzjahi:2003zy}.
This shift symmetry is broken by the $D$3-$\bar D$3 potential
$V_{D{\bar D}}(\phi)$, which is very weak due to warped geometry.
This inflaton potential is designed to break the shift symmetry
slightly, so inflation can end after slow-roll.

We shall study the wavefunction of tunneling to this type of 
string states. We shall find the maximum $\F$ for such inflationary 
vacuum as 
well as for the corresponding vacua without the extra pair of
$D$3-${\bar D}$3-brane needed for inflation. In each case, the fluxes
are determined by maximizing $\F$. If we know $\hat c$, this ansatz 
will select a specific stringy vacuum via quantum tunneling.
Since we do not know the value
of $\hat c$, we shall instead use the data to fix it.
Having fixed $\hat c$, we can then evaluate $\F$ for any vacuum.
Fortunately, the qualitative features are unchanged if we vary 
$\hat c$ by a few orders of magnitude. Before going into the discussions,
we give a brief summary of the key results.

\begin{itemize}

\item Using $G_N$ and the COBE density perturbation data, $M_s$ and 
$c$ can be fixed. We find $c \sim 10^{-3}$.
Maximizing $\F$ allows us to fix the values of the fluxes.

\item We find that tunneling to an KKLMMT-like inflationary universe 
(with ${\F} \sim 10^{18}$) is much preferred over the $S^{10}$ universe 
(with ${\F} \sim 10^{9}$) discussed earlier. 

\item This $D$3-${\bar D}$3 inflationary universe is also much
preferred over the KKLT vacua without inflation (${\F} < 0$). 
That is, direct tunneling to today's vacuum is strongly suppressed.

\item The KKLT string vacuum state that the inflationary universe 
ends in (after the $D$3-${\bar D}$3 branes have annihilated) is 
different from the vacuum that will be reached directly from 
tunneling. They have different fluxes.

\item More detailed analysis of the above model and investigation 
of other string vacua should help to narrow the uncertainty in 
$c$. A better understanding of $\Psi$ will allow us
to perform a more precise selection of a stringy vacuum state. 

\end{itemize}

\subsection{Set-up}

The moduli stabilization follows the approach due to 
GKP \cite{Giddings:2001yu} and KKLT \cite{Kachru:2003aw}, where
more background and details can be found. Here we give an overall picture
of the application to the wavefunction. Many details are saved for 
Appendix F.
The complex structure moduli are stabilized by the
introduction of quantized fluxes wrapping around some cycles while the
K\"{a}hler modulus is stabilized by a non-perturbative QCD-like effect. In
particular, let $M$ be the units of RR 3-form field strength in a 3-cycle
$A$ and $-K$ be the units of 
NS-NS 3-form field strength in the dual 3-cycle $B$. 
Here, $K$ and $M$ are discrete parameters. Different choices of $K$ and
$M$ correspond to different vacua. For fixed $K$ and $M$, the introduction
of an additional pair of $D3-{\bar D}3$-branes 
\cite{Kachru:2003sx}
raises the vacuum energy and
inflation takes place. It is argued \cite{Firouzjahi:2003zy}
that the slow-roll condition is generically satisfied so there will be
enough inflation before the pair of branes collides, annihilates, (re)heats
and starts the hot big bang epoch.

Start with the metric 
\ba
ds^2_{10}= e^{2u(y)} \eta_{\mu \nu} dx^{\mu}dx^{\nu}
+e^{-2u(y)} \hat g_{m n} dy^m dy^n
\ea
where $A(y)$ measures the initial size of the universe created via
tunneling. 
Consider the simple case of a single complex modulus $z$. There are
$2+2b_{2,1}=4$ 3-cycles, namely a pair $(A, B)$ (dual cycles that intersect
only once) and an additional pair $(A', B')$. 
The integrals are the periods defining the complex structure of the
conifold.
In particular, $z$ is the complex coordinate for the cycle $A$
\ba
z \equiv \int_A \Omega
\ea
where $\Omega$ is the $(3,0)$-form and that on the dual cycle is
\ba 
{\cal{G}}(z) \equiv \int_B \Omega = \frac{z}{2 \pi i} \ln z + 
\text{holomorphic}
\ea
The three-cycles A and B are in the vicinity of the Klebanov-Strassler 
conical point where fluxes are turned on:
\ba
\frac{1}{2\pi \alpha'}\int_{A}F_{(3)}=2\pi M \nonumber\\
\frac{1}{2\pi \alpha'}\int_{B}H_{(3)}=-2\pi K.
\ea
where $F_{(3)}$ is the 3-form field strength of the RR field 
and $H_{(3)}$ is the 3-form field strength of the NS-NS field.
Turning on $-K'$ units of $H_{(3)}$ on the $B'$ cycle, we finally obtain
the superpotential
\ba
\label{W}
W=&M {\cal{G}}(z)-K\tau z-K'\tau z'(z)+Ae^{ia\rho} 
\ea
where $\tau=C_{(0)}+ie^{-\phi}$ and the last term is a non-perturbative 
interaction contribution due to the presence of D7-branes.
Here, $z'$ is a function of $z$ which is generically non-vanishing at
$z=0$, $z'(0) \simeq 1$.

Following KKLMMT we have
\ba
\label{rhov}
\Lambda=8\pi G\, T_3 \frac{r_0^4}{R^4}(1-\frac{1}{N}\frac{r_0^4}{r_1^4})
\ea 
where $r_0$ and $r_1$ are the positions of the stack of $\bar{D}3$-brane 
and the $D3$-brane, respectively. $T_3$ is the D3-brane tension: 
\ba
\label{D3tension}
T_3=1/(2 \pi)^3 g_s \alpha'^2
\ea
and $R$ is the curvature radius of the AdS geometry:
\ba
\label{R}
R^4=4\pi g_s N \alpha'^2.
\ea
Here $N$ is the number of ${\bar D}3$-branes.
The $D3$ charge conservation requires
\ba
N=\frac{1}{2\kappa_{10}^2T_3}\int_{\cal{M}}H_{(3)}\wedge F_{(3)}=MK
\ea
To trust the low energy supergravity approximation, we assume $N\gg1$.
After compactification of the internal space we have
\ba
\label{Mpl}
M_{Pl}^2=\frac{2}{(2\pi)^{7}}\frac{V_6}{g_s ^2 \alpha'^4}
\ea 
where $M_{Pl}^{-2}\equiv8\pi G_N$ and
$V_6$ is the volume of the Calabi-Yau manifold. 
For simplicity we consider $V_6= \alpha'^{3} r^{6}$, where 
$r$ is related to the 
imaginary part of the K{\"a}hler modulus $\rho= b+ ir^4$. 

Considering first only the imaginary part of $\rho$, we find
\ba
\label{rhoG2}
-S_E =\frac{3\pi}{2G_N\Lambda}= \frac{6}{(2\pi)^9}\,
(\frac{\rho}{g_s})^{3}(\frac{R}{r_0})^4 \ .
\ea
Fluxes induce a large hierarchy of scale between $r_0$ and $R$: 
$(\frac{r_0}{R})^4=e^{-8\pi g_s K/3M}$.
We see that $S_E$ (\ref{rhoG2}) is exponentially large, due to the
warp factor $\frac{r_0}{R}$. For example, adopting the numbers 
in appendix C of KKLMMT, we find that $-S_E\sim 10^{21}$.
This implies that, order-of-magnitude wise, ${\F} \lesssim 10^{21}$.

\subsection{Applying the New Wavefunction}

Now we are ready to apply our ansatz to the tunneling from nothing to
the above inflationary vacuum in string theory.
Since the extra dimensions are stabilized already, we may assume that
the terms needed to suppress large compactification sizes are 
negligible. Keeping the leading term in $\cal{D}$, we have
${\F} \sim e^{-S_E - {\hat c}M_s^9 V_9}$, where $V_9$ is the spatial volume 
of the closed universe at the end of tunneling.

For $S^4 \times CY$, as long as the CY volume modulus and complex 
structures are 
fixed, we can use the effective four dimensional point of view such that 
$V_9=V_6 \times V_3$, where $V_6$ and $V_3$ are the spatial volumes of 
the CY manifold and our three dimensional space, respectively. We have
\ba
\label{V3cr}
V_3=\frac{2\pi}{3}\, (\frac{\Lambda}{3})^{-\frac{3}{2}} \ .
\ea
Using Eqs (\ref{rhov}), (\ref{D3tension}) and  (\ref{Mpl}) 
with the identity 
$V_6 \equiv  \alpha'^3 r^6 = \alpha'^3 \rho^{\frac{3}{2}}$, 
we find
\ba
\label{v9expression}
{\cal D} = {\hat c} M_s^9 V_9= {\hat c} \, \frac{2\sqrt{3}}{(2\pi)^5}\,
\rho^{\frac{15}{4}} g_s^{\frac{-3}{2}}(\frac{R}{r_0})^6 \ .
\ea
Maximizing $\Psi$ or $\F$ determines $M$ to be $M_1$ and $K$ to be $K_1$.
The details of the analysis can be found in Appendix F. For any given 
vacuum state, quantities like $z^{\prime}(0)$, $|{\cal{G}}(0)|$, $a$ 
and $A$ 
etc. are all in principle calculable. For our purpose here, we shall 
simply adopt some generic choices of values. 

In GKP it was assumed $|{\cal{G}}(0)|\sim 1$ and $z'(0)\sim 1$, which we 
shall adopt. In this approximation, we find, in Appendix F,
\ba
M_1 &\sim&  0.4 \times |A| \\\nonumber
K_1 &\sim& 0.8 \times \ln\left(10^{-4}a^{3/4}\, |A|^{3/2}\, 
{\hat c}^{-1} \right)
\\\nonumber
(\frac{r_0}{R})&\sim&115 \times a^{-3/8}\, |A|^{-3/4}\, {\hat c}^{1/2}\\\nonumber
\frac{M_{Pl}}{M_s}&\sim& 3\times 10^{-4}\,  a^{-3/4}\, |A|\\\nonumber
\frac{1}{g_s} &\sim& 0.1\times |A|\\\nonumber
r&\sim& (\frac{1.2}{a})^{1/4}
\ea
The value of ${\F}$ at the maximum is
\ba
{\F}_{max} \sim 2\times 10^{-18}\, a^{-3/2}\, |A|^6\, {\hat c}^{-2} \ .
\ea
or, expressing in terms of the COBE normalization, $\delta_{H}$:
\ba
{\F}_{max} \sim 1.6 \times 10^4 \, \delta_H^{-3} \ .
\ea 
For $\delta_H\sim 2\times 10^{-5}$, we find
\ba
{\F}_{max}\sim 10^{18} \ .
\ea
To trust the low energy supergravity approximation, we need $r>>1$ and 
$g_s<1$. 
The first condition requires $a<<1$. For example the value chosen
by KKLT, $a=0.1$, results in $r\sim 2$. One may choose even smaller value
of $a$ to improve the low energy supergravity approximation.
To satisfy the condition $g_s<1$, we simply choose $|A|>10$.

We see that all the physical outputs are functions of $a$, $ |A|$ and 
$\hat c$.
As an example, let us take $a=0.001$, $|A|=100$.
To get the COBE bound $\delta_H\sim 10^{-5}$, we find 
$${\hat c}\sim 10^{-10} \quad \rightarrow \quad c \sim 10^{-3}$$
With above values of $a$ and $|A|$, we get $M_1 \sim 40$ and $K_1 \sim 12$,
$\frac{r_0}{R}\sim 2\times 10^{-3}$ and $M_{Pl}\sim 6 \, M_s$. The closeness
of $M_s$ to $M_{Pl}$ may raise some concerns initially, but to find the actual 
physical scale of inflation one must take into account the warping effect. 
Including the effect of the warp factor, one finds
\ba
\rho_{v}^{\frac{1}{4}}\sim (\frac{r_0}{R})M_s \sim 10^{14}\,  GeV 
\ea
in agreement with the commonly used energy scale of inflation.
Summarizing the main results, we find that

\begin{enumerate}

\item  Using only the Hartle-Hawking wavefunction, the flux $M$ is 
fixed to a finite value (of order 10) while the flux $K$ is not determined.
 The larger is the value of
$K$, the larger is the tunneling probability. Since large $K$ exponentially
suppresses the vacuum energy, we end up with an universe with a very small
cosmological constant, as expected.

\item   Using our improved wavefunction, maximizing the tunneling 
probability to an inflationary vacuum determines both $M=M_1$ and 
$K=K_1$. For a reasonable choice of $c \sim 10^{-3}$, 
the preferred 
inflationary universe (that is, with the largest tunneling 
probability) easily fits the COBE data, with ${\F} \simeq 10^{18}$.

\item  Using the above value for $c$, the 10-dimensional de 
Sitter universe and similar vacua have ${\F} \simeq 10^{9}$. 

\item The KKLT vacua are very much like our today's universe.
It is easy to estimate the tunneling probability to such a vacuum state,
as done in Appendix F. We find that ${\F}_{KKLT} <0$, so they have 
extremely small tunneling probabilities. Taking $\Lambda$
to be today's dark energy value, we have ${\F} \sim - 10^{170}$.

\item To conclude, the universe prefers to go through an 
inflationary phase. 
This qualitative (order of magnitude)
result is largely independent of the details (see Figure \ref{fig3}). 
In this sense, we believe the picture is robust. 

\item  Let $K_0$ and $M_0$ be the values for the vacuum with the largest
tunneling probability to go directly to an universe bypassing an
inflationary phase. We find that $(M_1, K_1) \ne (M_0, K_0)$. This means
that vacua (such as the $(M_0, K_0)$ vacuum) without an inflationary 
road leading to them may be unreachable.

\item The above analysis is very crude. Note that $A$ and $a$ are
input parameters not 
determined by maximizing $\F$. Also, instead of order of magnitude 
estimates, 
functions like $z^{\prime}$, $\cal G$ etc. should be calculable. 
With a better wavefunction $\Psi$, we should be
able to find the preferred inflationary universe with no free parameters. 
It will be interesting 
to find $\F$ for other inflationary scenarios in string theory
\cite{Denef:2004dm,Burgess:2004kv}.

\item Although we find $c \sim 10^{-3}$, the analysis is quite crude. 
As is clear in the analysis, some 
reasonable variation of the parameters used can easily change $c$ by 
a few orders of magnitude. 
We believe $c \sim 1$ is quite possible.
This calls for a more careful study. 

\end{enumerate}

\section{Supercritical versus Critical String Vacua}

Although string theory is known to have 10 (11 for M theory)
as its critical dimension, higher-dimensional vacua are known to be possible
\cite{Chamseddine:1992qu,Silverstein:2001xn,Maloney:2002rr}.
In $D>10$ dimensions, the low energy effective theory containing the massless graviton, 
dilaton and R-R fields is given by (in Einstein frame) \cite{Chamseddine:1992qu}
\ba
\label{generalDaction}
S_D=\frac{1}{{2\kappa_D}^2}\int d^{D}\, x \sqrt{-g}\left( R-\frac{2(D-10)}{3\alpha'}e^{4\phi/(D-2)}\,\,\,\,\,\right.\nonumber\\
-\left.\frac{4}{D-2}\nabla_{\mu}\phi \nabla^{\mu}\phi-\frac{1}{2}\sum_{p}e^{4(1-p)\phi/D-2}(F_p)^2\right)
\ea
where the sum runs over various RR fields in the theory, $\phi$ is the dilaton and 
$\kappa_D$ stands for D-dimensional gravitational constant:
\ba
\label{kappaD}
2\kappa_D^2\equiv 16\pi G_D =g_s^2(2\pi)^{D-3}M_s^{2-D} \equiv 
g_s^2l_s^{D-2}/2 \pi
\ea
Note the presence of the cosmological constant term, due to the linear dilaton background.
This term is absent in $D=10$, allowing $D=10$ flat spacetime.
For $D>10$, $4$-dimensional flat, or almost flat, spacetime may be achieved in the 
presence of contributions from the many RR fluxes, branes and orientifold planes.
In general there are many local dS vacua along with a global minimum, corresponding 
to $\phi \rightarrow -\infty$.
The resulting cancellation can in principle yield an almost stable vacuum 
with a  parametrically small cosmological constant that attains the value of the observed 
dark energy \cite{Silverstein:2001xn,Maloney:2002rr}.

Such supercritical vacua in general do not have supersymmetry. So the 
liklihood of low energy supersymmetry, a very important phenomenological issue,
hinges crucially on which type of vacua 
nature prefers, critical or supercritical.
We like to check if SOUP shed light on this fundamental question.
A general study in SOUP program to find the dimensionality of the space-time is rather 
delicate and it may need a good knowledge of detailed properties of each vacuum.
However, as a first test to illustrate the proposal, let us consider
the D-dimensional instanton $S^D$ for $D>10$. The main contribution for the 
cosmological constant is given above
\ba
\label{lambdaD}
\Lambda(D)=\frac{2(D-10)}{3\alpha'} g_s^{\frac{4}{D-2}}
\ea 
We would like to calculate ${\F}$ for different $S^D$ for $D>10$, and compare it
to the value of ${\F}$ for $S^{10}$, the critical case.
 
Using Eq.(\ref{area}) for the surface area of a unit $S^D$ instanton along with 
Eqs.(\ref{kappaD},\ref{lambdaD}), we find
\ba
\label{generalDF}
{\F}=4 \pi^2\sqrt{\pi}g_s^{-2}\left(\frac{3}{2(D-10)}\right)^{\frac{D-2}{2}}
\left(\frac{(D-1)(D-2)}{8\pi}\right)^{\frac{D-1}{2}}\nonumber\\
\times
\left[ 4\sqrt{\frac{\pi(D-1)}{2(D-2)}}\frac{1}{\Gamma(\frac{D+1}{2})}
-\frac{c}{(2\pi)^2g_s^{(D-2)/2}}\sqrt{\frac{3}{2(D-10)}}\frac{1}{\Gamma(\frac{D}{2})} \right] \ .
\ea
Using $c \sim 10^{-3}$ and $g_s=1$
we find that ${\F}$ quickly goes to zero for large $D$:
(${\F} \simeq 10^{9}$ for $D = 10$, ${\F} \simeq 10^{4}$ for $D = 11$, 
${\F} \simeq 10^{3}$ for $D=12$
, ${\F} \simeq 12$ for $D=16$ and ${\F} \simeq 0.7$ for $D=20$.) 
This is consistent with the general expectation that for large $\Lambda$, 
${\F}\rightarrow 0$, as is indicated in Figure {\ref{fig3}}.
Adding a non-BPS brane or brane pair will simply decrease $\F$ further.
As we increase $D$, the semi-classical approximation eventually breaks down, but 
we do not expect any change in the qualitative picture.   
Interestingly enough, we see that ${\F}$ for $S^D$, $D>10$ is negligible 
compared to the value ${\F}\sim 10^{9}$ for $S^{10}$, or to the KKLMMT
inflationary model with ${\F}\sim 10^{18}$. Naively, we may 
conclude that critical string vacua are preferred over supercritical string vacua. Unfortunately, such a conclusion is premature.

It remains to be seen if some sort of brane inflationary scenario may be implemented 
in the supercritical string framework. This is a big if. Suppose a brane inflationary 
scenario similar to that discussed earlier can be implemented. Let us further assume 
that both inflationary scenarios (in critical and in supercritical strings)
that agree with our early universe are among the vacua preferred by SOUP. Can SOUP then 
distinguish between them ? In this case, it turns out that higher correction terms 
to $\F$ is 
required to distinguish them. To see why, let us start with
$G_N = G_D/V_{D-4}$,
where $V_{D-4}$ is the volume of the compactified dimensions.
Using Eq.(\ref{V3cr}), a simple generalization of the earlier discussions gives
\be  
\label{sup}                                                   
{\F}=\frac{3\pi V_{D-4}}{2G_D \Lambda}- c \frac{2\pi V_{D-4}}{3({\Lambda}/3)^{3/2}
l_s^{D-1}}
= \frac{3\pi }{2G_N \Lambda} - 
c \frac{g_s^2}{48 \pi G_Nl_s({\Lambda}/3)^{3/2}}
\ee
For $D$-independent $c$, as suggested by the decoherence argument, 
there is no explicit $D$-dependence in $\F$. With $c \sim 10^{-3}$, 
maximizing $\F$ yields $\sqrt{\Lambda} \sim 10^{14}$ GeV and 
${\F} \sim 10^{18}$, independent of $D$. To distinguish them, we may 
need a better knowledge of $\F$. It is also important to check if a 
viable inflationary scenario may be incorporated into supercritical 
string vacua. In that case, maximizing $\F$ and fitting data
will determine $g_s$ and $\Lambda$. We can then compare 
the probabilities of tunneling to such a supercritical inflationary 
scenario and to the most favorable critical inflationary scenario.

For example, fixing $l_s$, $c$ and setting 
$\partial {\F}/\partial \Lambda =0$ 
yields ${\F} \propto g_s^{-4}$.
With explicit supercritical inflationary scenarios,
SOUP may be able to express its preference between critical and 
supercritical inflationary vacua.
If, on the other hand, it is $\hat c$ instead of $c$ that is 
$D$-independent, then
${\F} \propto g_s^{-4}(2 \pi)^{-2(D-10)}$. In this case, critical 
string vacua are likely to be preferred. It is clear that further 
study will be important to find the preferred vacuum.

\section{Remarks} 

In this paper, we propose an improvement of the Hartle-Hawking
wavefunction. The key difference in the new wavefunction is the
inclusion of the 
the backreaction effects due to the interaction of the metric with 
the matter fields. These matter fields (multipole modes perturbing the
background classical metric as well as other bosonic and fermionic 
modes) play the role of the environment and are traced over, providing 
a decoherence term that tends to suppress the tunneling to
universes with small cosmological constants. As a consequence,
intermediate values of the vacuum energy seems to be preferred.  
This is precisely what we want, since this allows the tunneling to 
an inflationary universe not unlike the one our universe has gone through.
To conclude, let us make some remarks :

\begin{itemize}

\item

Here we like to point out that the above proposal should be further 
modified. Consider an instanton that leads to a universe with little 
or no inflation. That universe will either recollapse, 
or reach only a very small size.
As a result, the likelihood for us to be inside such a region will 
be extremely small compared to a universe that goes through 
an extended period of inflation. As a simple ansatz, the likelihood 
$P_S$ of an observer ending in a particular universe $S$ with size 
$V_S$ should be
\ba
P_S= \frac{V_S}{V_I}|{\cal P}|^2 e^{2{\F}}
\ea
where $V_I$ is the volume of the initial universe just after 
tunneling, and $V_S$ is the volume immediately after 
inflation. Here $\cal P$ is the prefactor that is calculable in principle.

\item Following the Hartle-Hawking prescription, it seems that 
a universe with an arbitrarily small cosmological constant is 
preferred. This corresponds to an arbitrarily large instanton, 
which corresponds to infinite (Euclidean) time for the
``nucleation'' bubble creation. 
This is clearly unacceptable, because it implies that the universe 
as big as today's universe is much more likely to be created directly 
from nothing than a much smaller universe, say our universe 13 
billions years ago. We argue that there must exist a 
mechanism that suppresses this effect, such that a vacuum energy density 
acceptable for inflation is naturally selected. 
Including the coupling of matter fields and higher gravitational modes, 
we are led to a modified wavefunction, which includes a decoherence term.
It is important to give a more complete argument for the 
wavefunction we proposed, and if additional modifications should be 
included.
It will be a challenge to calculate the value of the coefficient 
$c$. 

\item 
String theory has many vacua/state with very different vacuum energies.
We use the proposed wavefunction to compare the probability of tunneling 
from nothing to various geometries, albeit a limited set.
By comparing the tunneling probabilities for the different geometries,
we see that tunneling from nothing to an inflationary universe 
very close to what our universe went through is favored.
That is, tunneling directly to a universe with a much smaller 
cosmological constant (like today's dark energy) is very much suppressed.
Tunneling directly to a 10-dimensional de Sitter universe is also 
suppressed.

\item Not all viable string vacua allow an extended period of inflation
that ends with a hot big bang. Only those string vacua that are sitting 
at the end of a long inflationary road have any chance of being our vacuum. 
That is, we will not end up in a vacuum which is not close to a
favorable inflationary road.
In the search of why our particular vacuum is selected, this 
should cut down substantially the number of string vacua one has 
to take into consideration.

\item So far, we have considered only a very limited set of geometries.
It will be important to check if a universe with $3$ inflating 
dimensions is really preferred over all other geometries. For example,
tunneling to a Randall-Sundrum brane world (i.e., a Horava-Witten model) 
\cite{Garriga:1999bq,Hawking:2000kj} is worth a new analysis within SOUP.
Clearly more investigation is required with realistic string models 
to see if our particular $4$-dimensional universe is favored 
over other $4$-dimensional universes. 

\item For SOUP to be phenomenologically meaningful, we must use the 
Hartle-Hawking wavefunction and not the alternative proposed in 
Ref.\cite{Linde:1984mx,Vilenkin:1985dy}. In practical terms, one may view 
this as a resolution of the 20 year old debate.

\item Since it is clear that the string theory has many instanton solutions,
we must entertain possibility of a collection of instantons each of 
which gives rise to its respective universe. 
This also suggests that, semiclassically
the wavefunction should include the sum over the 
metric that includes disconnected pieces.
This must allow the collision of different universes.
In terms of string theory, presumably $\Psi$ is a 
wavefunction in string field theory. Semiclassically, one may treat
closed string modes, in particular gravity, as background fields 
with small fluctuations. Such fluctuations allow topology changes
and so the interaction among the topologically distinct universes
should be included.
This is clearly interesting to investigate further.

\item $S_{entropy}=-S_E$ is the entropy of de Sitter space, maximizing 
the tunneling probability with the Hartle-Hawking wavefunction
is equivalent to maximizing this entropy.
Recall the von Neumann entropy expressed as a function of the density 
matrix, 
\be
S_{vN} = - Tr(\rho \ln \rho) 
\ee
Going from pure-state to mixed-state increases $S_{vN}$. It is tempting
to relate the maximizing of the tunneling probability with the modified
wavefunction to the maximizing of the von Neumann entropy $S_{vN}$ for 
the tunneling process.
 
\item It is possible that such a SOUP cannot be derived from string 
theory as we know it, since the derivation of the wavefunction 
requires a knowledge of the full wavefunction $\Psi(a=0,x_n)$ of 
nothing (i.e., no classical spacetime). This means that SOUP must be 
postulated. If so, it is even more important to take a 
phenomenological approach to learn about SOUP. In this case, it is
unclear how to differentiate SOUP from the anthropic principle. 
One may view this as an attempt to quantify the anthropic principle.
In this paper, we have taken a more optimistic view. Since classical 
spacetime are merely derived quantities in string theory, as opposed to 
the case in general relativity, $\Psi(a=0,x_n)$ should be a meaningful
quantity in string theory, although a non-perturbative description
may be involved.

\end{itemize}
\vspace{0.5cm}

{\large{\bf{Acknowledgments}}}\\

We thank Andy Albrecht, Dave Chernoff, Alan Guth, Thomas Hertog, 
Nick Jones, Bjorn Lange, Andrei Linde, Joe Polchinski, Jim Hartle, 
Jeff Harvey, Don Marolf, Rob Myers, Jim Sethna,
Eva Silverstein and Alex Vilenkin
for useful discussions. This material is based
upon work supported by the National Science Foundation under Grant
No.~PHY-0098631.
\vspace{0.5cm}


\appendix

\section{Tunneling Amplitude}

The Feynman functional integral is over all 4-geometries with a spacelike 
boundary on which the induced metric is $h_{ij}$ and which to 
the past of that surface there is nothing (see Figure 1). 
$\Psi[h_{ij}]$ satisfies the Wheeler-DeWitt equation :
\be
{\cal{H}} \Psi = \left(-G_{ijkl} \frac{\delta^2}{\delta h_{ij} \delta h_{kl}}
- {^3R}(h)h^{1/2} +2 \Lambda h^{1/2} \right) \Psi[h_{ij}]=0
\ee
where $G_{ijkl}$ is the metric in superspace:
\be
G_{ijkl}=\frac{1}{2 h^{1/2}}( h_{ik}h_{jl} +h_{il}h_{jk}-h_{ij}h_{kl})
\ee
where $^3R$ is the scalar curvature of the intrinsic geometry 
of the 3-surface. For the de Sitter metric, ${\cal{H}}$ reduces to,
up to a constant factor,
\be
{\cal{H}} = \frac{1}{2} \left(-\frac{p^2}{a} - a +H^2a^3 \right)
\ee
where $p={\dot a}a$.
Upon quantization, $p \rightarrow id/da$, and the 
Wheeler-DeWitt equation becomes
\be
\label{pot}
\left(\frac{-d^2}{da^2} + U(a) \right) \Psi(a)=0   \quad U(a)=a^2(1-H^2a^2)
\ee
which takes the form of a 1-dimensional Schroedinger equation 
for a particle described by the coordinate $a(t)$.
This is illustrated in Figure 6. Adding matter fields $\chi$ gives
the equation:
\be
\label{wd}
\left( \frac{-\partial^2}{\partial a^2} + U(a) + \frac{\partial^2 \chi}{\partial \chi^2} - \chi^{2} -2\epsilon_{o}\right) \Psi(a, \chi) = 0
\ee
where $\epsilon_{o}$ comes from the possibility of a matter-energy 
renormalization. Let 
\be
\Psi(a, \chi) = \Sigma_{n} \psi_{n}(a) u_{n}(\chi)
\ee
then the above equation becomes two equations:
\baray
\label{two}
\frac{1}{2}\left(\frac{-d^2}{d\chi^2} + \chi^2 \right) u_{n}(\chi)
= (n + 1/2)u_{n}(\chi) \\ \nonumber
\frac{1}{2}\left( \frac{-d^2\psi_{n}}{da^2} + U(a)\psi_{n}\right)
= \epsilon \psi_{n} 
\earay
where $\epsilon = n + 1/2 - \epsilon_{0}$. For $\epsilon \to 0$ the above 
equation for $\psi_{0}$ just becomes Eq.(\ref{pot}) which describes 
the ground state of the universe. For nonzero $\epsilon$
one obtains the excited states of the universe.

Ignoring the matter fields, this reduces to Eq.(\ref{pot}). The 
classically allowed region is $a \ge 1/H$, and its solutions are :
\be
\Psi_{\pm}(a) \simeq e^{\pm i \int^a_{1/H} p(a') da' \mp i \pi/4}
\ee
In the classically forbidden region, i.e., the under-barrier $a < 1/H$
region, the solutions are
\be
\tilde \Psi_{\pm}(a) \simeq e^{\pm \int_a^{1/H} |p(a')| da'}
\ee
The under-barrier wavefunction is a linear combination of 
$\tilde \Psi_{+}(a)$ and $\tilde \Psi_{-}(a)$, where $\tilde \Psi_{+}(a)$
decreases exponentially  while $\tilde \Psi_{-}(a)$ grows exponentially
with increasing $a$. There are a number of proposals in obtaining the 
tunneling probability from this wavefunction.
\begin{figure}
\begin{center}
\epsfig{file=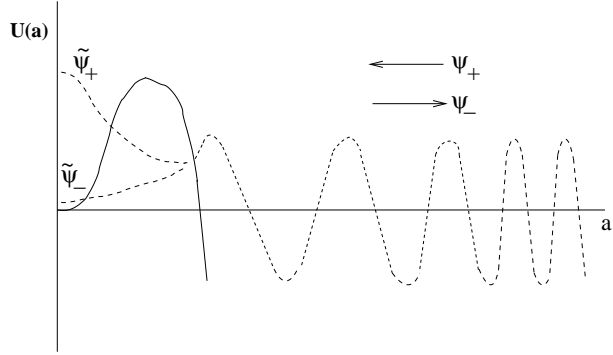, width=8cm}
\vspace{0.1cm}
\caption{The tunneling wavefunction. The growing and the decaying 
components under the barrier combine to form the outgoing 
and incoming waves on the right.
If only the growing mode is present (the Hartle-Hawking wavefunction),
or if only the decaying mode is present (the Linde wavefunction),
there will be both outgoing and incoming waves on the right. 
Vilenkin demands a particular combination so there is only outgoing wave.
Decoherence will yield the outgoing wave as the classical solution. 
}
\label{wavef}
\end{center}
\end{figure}
For quantum tunneling, the Hartle-Hawking wavefunction is given by
\be
\Psi[h_{ij}] = \int_{\emptyset}^{h_{ij}} [dg] e^{-S_E[g]} \simeq e^{-S_E}
\ee
where $S_E$ is the Euclidean action of the corresponding
instanton solution. This corresponds to the simple extension 
from quantum field theory. However, in gravity, $S_E$ is 
unbounded from below. In fact, for the de Sitter instanton solution,
one has Eq.(\ref{HHWF1}).
In terms of the under-barrier wavefunction, this corresponds to 
the dominance of the growing $\tilde \Psi_{-}(a)$. Our work is built 
on this interpretation.

According to DeWitt \cite{Dewitt:1967yk}
, the wavefunction at $a=0$ should vanish:
$\Psi_{DW}(a=0)=0$.
DeWitt believes that $a=0$ corresponds to a cosmological singularity,
so his condition avoids the occurrence of the singularity.
This implies that $\tilde \Psi_{+}(a)$ and $\tilde \Psi_{-}(a)$ cancels
each other at $a=0$. Since they are comparable at $a=0$, 
$\tilde \Psi_{-}(a)$ dominates around $a=1/H$. This gives the
same result as the Hartle-Hawking wavefunction.
This same result is also obtained if one assumes a generic 
wavefunction around $a=0$, so that the 2 
wavefunctions $\tilde \Psi_{+}(a)$ and $\tilde \Psi_{-}(a)$ 
are comparable around $a=0$.

If one assumes that only the decaying wavefunction is present at 
$a \sim 0$, then one obtains the following wavefunction,
$\Psi_L(a) \simeq e^{+S_E}$, as proposed by Linde \cite{Linde:1984mx}.
According to Vilenkin \cite{Vilenkin:1985dy,Vilenkin:1986cy}, 
$\Psi_-(a)$ corresponds to an expanding 
universe while $\Psi_+(a)$ corresponds to a contracting universe.
Suppose one wants only an expanding universe in the classically 
allowed region $a > 1/H$, that is, $\Psi(a > 1/H) = \Psi_-(a)$.
Then $\tilde \Psi_{+}(a)$ and $\tilde \Psi_{-}(a)$ should 
match around $a = 1/H$, or
\be
\Psi(a < 1/H) = \tilde \Psi_{+}(a) -\frac{i}{2} \tilde \Psi_{-}(a)
\ee
In this case, the under-barrier region is dominated by
$\tilde \Psi_{+}(a)$ and the tunneling wavefunction is
$\Psi_T \sim e^{-3 \pi/2G_N\Lambda}$.
Although this agrees with $\Psi_L$ for the de Sitter instanton, 
it is different in that
\be
\Psi_T = \int [dg] e^{iS} \rightarrow e^{-|S_E|}
\ee
so for ordinary quantum field theories with bounded Euclidean action, it 
reduces to the answer we know. This alternative possibility was 
also mentioned in Ref.\cite{Hartle:1983ai}.
However, as pointed out by Rubakov \cite{Rubakov:1984bh} and others, 
the arrow of time is not predetermined since there is no outside 
observer. Furthermore, the introduction of scalar fields into the 
gravity theory will lead to catastrophic radiation of scalar fields
(due to the sign flip of the matter field effective action). 
For our purpose, this proposal will lead to the breakdown of either 
the semiclassical approximation or to the unboundedness of the
string vacuum. As pointed out in Ref\cite{Halliwell:1989vw}, 
decoherence will
select the growing mode as the classical evolution of the universe.

The Hartle-Hawking wavefunction implies that quantum tunneling prefers
a de Sitter universe with size $a \simeq 1/\sqrt{\Lambda} \rightarrow
\infty$, which is clearly unphysical. This is a major reason in the 
search of an alternative wavefunction. Here we shall start with 
the Hartle-Hawking wavefunction. As discussed in the text, 
this wavefunction needs an important improvement due to the 
coupling/presence of other modes/fields.

\section{Feynman Paths with Complex Metric}

Here we follow closely the formalism of 
Halliwell and Louko \cite{Halliwell:1989vu}. Their normalization 
is slightly different from that used in the text. 
The metric ansatz is
\ba
ds^2=\sigma^2 \left(N(\tau)^2d\tau^2+b(\tau)^2 d {\Omega}_3^2 \right)\ ,
\ea
where $\sigma^2=2G/3\pi$. We choose $\tau$ to be real and a
(piece-wise) constant lapse function: $\dot N=0$. 


The Einstein-Hilbert contribution to the action (\ref{action4}) is
\ba
\label{SEH}
\frac{1}{16 \pi G} \int_M d^4x \sqrt{|g|} \left(R - 2 \Lambda \right)
=\frac{1}{2}\int d\tau N(-\frac{b {\dot b}^2}{N}+\lambda b^3-b)
+\frac{1}{2}\frac{b^2 \dot b}{N}{\large |}_{\tau=1} \ ,
\ea
where $\lambda=\sigma^2 \Lambda /3=\sigma^2 H^2$ 
and the last term above comes 
from partial integration of a second derivative term in the Ricci 
scalar, containing $\ddot b$. 
The extrinsic curvature is 
\ba
\label{extrinsicK}
K=-\frac{3\dot b}{\sigma N b} \ .
\ea
so the boundary term contribution to the action (\ref{action4}) is
\ba
\label{Sboundary}
\frac{-1}{8 \pi G} \int_{\partial M} d^3x \sqrt{|h|} K 
=-\frac{1}{2}\frac{b^2 \dot b}{N}|_{\tau=1} \ .
\ea
and
\ba
\label{totalS}
S=\frac{1}{2}\int^1_0 d\tau N(-\frac{b {\dot b}^2}{N}+\lambda b^3-b)\ . 
\ea
Taking the initial value of $b(0)=0$ and the final value $b(1)= a$, 
the solution of the action is
\ba
\label{b}
b(\tau)=r\sin(\frac{N\tau}{r}) \ .
\ea
where $r=1/\sqrt{\lambda}$. However, we shall leave $r$ unfixed, 
since we are interested in summing over four-spheres of 
arbitrary radius $r$. Instead of $r$, one may trade it for $z$, where
\ba
z\equiv1+\frac{\dot b(1)}{N}=1+\cos(\frac{N}{r})
\ea
and
\ba
r^2=\frac{a^2}{z(2-z)}
\ea
Inserting this ansatz in the action (\ref{totalS}) and performing the 
the $\tau$ integral we find
\ba
\label{complexS}
S(z,a,T)=S(z,a)=\frac{a^2}{6}\left(1-z+\frac{\lambda a^2-4}{z}+
\frac{\lambda a^2}{z^2}\right) \ ,
\ea
and the path integral is just a single ordinary integral over $z$:
\ba
\Psi(a)=\int_C dw dz \mu(z,a,w)\exp(-S(z,a,w))
\ea
Note that we have introduced $w$ to label the paths that have the same 
degenerate action (\ref{complexS}), a fact that will be important later.

We shall adopt the steepest-descent method, where the 
steepest-descent paths are the paths with $Im(S)=$constant. 
For $\lambda a_1^2 <1$, the saddle points are
\ba
z=1\pm(1-\lambda a_1^2)^{\frac{1}{2}} \ ,
\ea
which correspond to $r=\lambda^{-1/2}$ and represent, respectively, 
less than or more than half a four-sphere with three-boundary 
with radius $a$. (We ignore another saddle point at $z=-2$, 
which apparently corresponds to negative $\Lambda$.) 
The action at these two saddle points has the values
\ba
S(a_1)=-\frac{1}{3\lambda}\left(1\mp (1-\lambda a_1^2)^{\frac{3}{2}} \right) \ .
\ea
respectively. One may envision that $N$ stays real as the 
tunneling from $b(0)=0$ to $b(1)=a_1$, at which point the universe 
starts evolving classically in an inflationary epoch, where $a$ 
grow from $a_1$ to $\hat a$. That is, at 
$\tau=1$, we switch from the Euclidean metric (for $S^4$) to a 
Lorentzian metric (for de Sitter space), i.e., the 
lapse function $N$ goes from a real constant to a pure imaginary 
constant. To describe the tunneling of an inflationary universe 
from nothing, we must deal with values of $N$ in the complex $N$ plane.

Once we are ready to entertain complex $N$, let us consider the 
the tunneling from nothing directly to an universe with size $\hat a$,
where $\lambda {\hat a}^2 >1$.
In this case, the path integral has
the saddle points at (again ignoring the saddle point $z=-2$)
\ba
z=1\pm i(\lambda {\hat a}^2 - 1)^{\frac{1}{2}} \ ,
\ea
at which the action has the value
\ba
\label{SDpS2}
S(\hat a)=-\frac{1}{3\lambda}
\left( 1 \mp i(\lambda {\hat a}^2-1)^{\frac{3}{2}} \right) \ .
\ea
respectively. Here, the saddle points 
still correspond to $r=\lambda^{-1/2}$, but $r< {\hat a}$. 
We still have $b(0)=0$, $b(1)=\hat a$ and $\dot b(0)/N=1$, but
\ba
\label{complexpath}
N&=&r \left[ \frac{\pi}{2} \mp i u \right] \\\nonumber
a(\tau)&=& r\sin \left(\frac{\pi \tau}{2} \right) \cosh(u\tau) 
\mp i \cos \left(\frac{\pi \tau}{2}\right) \sinh(u\tau) \\\nonumber
\cosh(u)&=& \lambda^{1/2}\hat a
\ea
These new solutions arise only when one 
performs the analysis with complex metric.

Now, let us come to the issue of degeneracy.
Introducing $T=N\tau$, the action (\ref{totalS}) becomes
\ba
S=\frac{1}{2} \int ^N_0 dT 
\left[ -b \left(\frac{db}{dT}\right)^2 +\lambda b^3 -b \right]  
\ea
this is a complex integral over $T$ of an analytic function 
in which the contour in the complex $T$ plane can take a variety 
of paths, a subset of which is shown in Figure 5.

It is instructive to calculate the extrinsic curvature. Using 
Eqs (\ref{extrinsicK}) and (\ref{b}) we find
\ba
K=-\frac{3\sqrt{\lambda}}{\sigma}\, 
\text{cotan}\left(\tau(\frac{\pi}{2}\mp iu)\right) \ .
\ea
For $\tau$ in the interval $0\le \tau \le 1$, $K$ is in general a 
complex number. At $\tau=1$ we find
\ba
K=-\frac{3i}{\sigma\hat{a}}(\lambda {\hat{a}}^2-1)^{\frac{1}{2}} .
\ea

\section{Decoherence}

Decoherence introduced in Eq.(\ref{master}) is brought about by lifting 
the degeneracy of complex metric paths by introducing matter degrees
of freedom. However, the concept of decoherence is relevant and 
applicable in a broader framework. Decoherence can be understood in 
very general terms in quantum mechanics as arising due to the 
interaction between the system and the environment. 
The effect of decoherence has been explored in the literature in 
various quantum mechanical situations, including quantum cosmology. 
In any quantum mechanical problem there is  a system, consisting
of the part that we are interested in studying, and the environment,
that consists of the part that is not directly relevant. The environment
can interact with the system in such a way that leads to a loss of
quantum mechanical features (i.e., a loss of coherence) and this process
is called decoherence. This may be compared to the quantum Zeno effect 
where a continuous measurement of the system by the environment leads 
to the suppression of its tunneling \cite{Chiu:1977ds,Peres:1980ux}. 

A simple example consists of an electron beam (the system) in the double 
slit experiment in quantum mechanics with a gas of molecules between 
the slits and the screen (the environment). The random 
interactions between the gas molecules and the electrons can lead 
to a loss of coherence and the interference pattern can get washed 
out leading to a decoherent pattern on the screen. The idea of 
decoherence has also been applied to quantum cosmology to study the 
semi-classical evolution of the universe. 
Although one would expect the universe to evolve as some superposition 
of various possible histories (for example, a quantum mechanical 
superposition of the growing and the decaying mode in the case of 
the de Sitter solution), decoherence allows the universe to evolve 
classically. 

The decoherence term in Eq.(\ref{master}) ought to arise as a result 
of such interaction between the system (the gravitational instanton)
and the environment (consisting of all other fields) 
\cite{Kiefer:1987ft,Halliwell:1989vw,Kiefer:1989ud}. 
Decoherence should forbid the quantum tunneling of a large 
(macroscopic) sized universe. 
Whereas in previous works \cite{Kiefer:1987ft,Halliwell:1989vw}
the idea of decoherence has been applied   
to the outside barrier region of the potential in Eq.(\ref{pot}) to study
the classical evolution of the universe, we attribute the origin of
the decoherence term in Eq.(\ref{master}) to the decoherence mechanism
in the under barrier region. So it is an extension of the previous ideas
to the under barrier region. 

It is convenient to use the language of density matrix to study decoherence.
It is also possible to cast the arguments using path integral language but
we shall use the density matrix language here. Consider the total 
wavefunction  $|\Psi>$ of the form :
\be
|\Psi> = \sum_{n}c_{n}|S_{n}>|E_{n}>
\ee
The corresponding pure-state density matrix is:
\be
\rho = |\Psi><\Psi|=\sum_{mn}c_{n}c^{*}_{m}|S_{n}>|E_{n}><E_m|<S_{m}|
\ee
where $|S_{n}>$ are states of the system and $|E_{n}>$ are states of the
environment. In this state, the system and the environment are correlated 
with each other. We are, however, only interested in the state of the 
system, not the environment. So the environmental degrees of freedom 
can be traced over 
in any calculation of interest. The object of interest is then the reduced 
density matrix given by:
\be
\rho_{S} = tr_{E}|\Psi><\Psi| = \sum_{n,m} c_{n}c_{m}^{*}<E_{m}|E_{n}> 
|S_{n}><S_{m}|
\ee
If the dot products $<E_{m}|E_{n}>=\delta_{mn}$,
the off-diagonal terms of the density matrix vanish
so there are no interference terms, and the system 
displays a pure classical behavior. 
The reduced density matrix then becomes diagonal:
\be
\rho_{S} = \sum_{n} |c_{n}|^{2}|S_{n}><S_{n}|
\ee
In general, $<E_{m}|E_{n}>$ is only suppressed when $m \neq n$.
Decoherence is the suppression of the off-diagonal terms, so the system
behaves almost classically.
In quantum cosmology the total wavefunction is
\be
\Psi (a, {x_{n}})=  \psi_{o}(a)\prod_{n>0}\chi_{n}(a,x_{n})
\ee
where $\psi_{o}(a)$ is the Hartle-Hawking wavefunction, and 
$\chi_{n}(a,x_{n})$ are the contributions from all other fields 
with amplitudes $x_{n}$. These
fields interact with the gravitational contribution encoded in $\psi_{o}(a)$
and the effect of this interaction can be seen 
by considering the reduced density matrix:
\be
\rho(a ; a^{'} ) = tr_{x_{n}} \left( |\Psi><\Psi| \right) \\ \nonumber
= \psi_{o}(a)\psi_{o}^{*}(a^{'}) \prod_{n>0} \int_{-\infty}^{\infty}
dx_{n}\chi_{n}(a, x_{n})\chi_{n}^{*}(a^{'}, x_{n})
\ee
Here $\rho(a ; a^{'} )$ should be compared with the density matrix 
element $\rho_{mn}$. In the absence of the environmental degrees of 
freedom, the density matrix
element corresponding to the tunneling of the universe as derived from
the Hartle-Hawking wavefunction is given by $\rho(a = H^{-1};0)= e^{-S_E}$.
However, tracing out the environment is expected to give the 
decoherence term leading to the wavefunction proposed in Section $2$,
\be
\Psi(a)=\rho(a = H^{-1} ; 0 ) \simeq e^{-S_E - \cal{D}}
\ee
The exact form of $\cal{D}$ depends on the particular model that we 
consider. A calculation can be performed along the lines in 
Refs \cite{Kiefer:1987ft,Halliwell:1989vw} with
the environment given by the higher multipoles of the matter and geometry
as obtained by Halliwell and Hawking \cite{Halliwell:1985eu}. 
That tracing out these  multipoles leads to decoherence in the 
Lorentzian evolution of the 
universe has been shown in \cite{Kiefer:1987ft}. 
We are proposing that a similar procedure
can be applied to the under barrier Euclidean/complex metric 
regime and  that can lead to the 
appearance of the decoherence term $\cal{D}$. In this sense, the 
inclusion of the complex
metric paths and the $S^{'}$-brane is related to the appearance of 
decoherence by tracing out the environment. The decoherence
should appear in string theory by tracing over the closed and 
open string degrees of freedom. Hopefully, string theory provides a 
natural cut-off that is absent in the analysis so far.

 Although the exact form of  $\cal{D}$ is expected to be obtainable only
from a complete analysis in string theory, we shall give a heuristic
derivation here which shall justify the form of $\cal{D}$ we have 
claimed. Consider a geometry of the type $S^4 \times M$ where 
$M$ is a $6$-dimensional compactified space of volume $V_{6}$. Consider 
only the tensor modes as the environmental degrees of freedom as in 
\cite{Kiefer:1987ft}. To trace over the environment, we need to 
know the wavefunction of these tensor modes in the under-the-barrier 
Euclidean region. However, as shown in Ref.\cite{Halliwell:1985eu}, the 
form of the tensor mode
wavefunctions is the same both for the Euclidean and the Lorentzian regions
and this is especially true for large $n$ (small wavelengths) modes.
In particular, for both the Euclidean and the Lorentzian cases, the
tensor modes are given by:
\baray
\label{pert}
\chi_{n}(a, x_{n}) = \left( \frac{n a^{2}}{\pi}\right)^{1/4} \exp \left( -2i \frac{\partial S}{\partial \alpha}x_{n}^{2}- \frac{1}{2}n e^{2 \alpha}x_{n}^{2}  \right )
\earay
where $S$ is the Hamilton-Jacobi parameter. The phase parts differ for the 
Euclidean and the Lorentzian regions. But it is the amplitude part of the
tensor modes that are important for determining the decoherence term and 
they are the same, viz. $\left( \frac{n a^{2}}{\pi}\right)^{1/4} \exp \left(- \frac{1}{2}n e^{2 \alpha}x_{n}^{2}  \right )$ for both regions. 
Tracing over the tensor modes involves finding
\baray
\int dx_{n}\chi_{n}^{*}(a, x_{n})\chi_{n}(a^{'}, x_{n})  = \sqrt{2}\left( \frac{a^{2} + a^{'2}}{a a^{'}}\right)^{-1/2}
\earay
This leads to the result we have used in Eq.(\ref{den})
\be
\rho_S(a ; a^{\prime}) \propto 
\prod_{n>0}
\exp \left(- \frac{(a+a^{\prime})^2(a-a^{\prime})^2}
{4a^2a^{{\prime} 2}} \right) \sim 
\exp \left(- N\frac{(a+a^{\prime})^2(a-a^{\prime})^2}
{4a^2a^{{\prime} 2}} \right)
\ee
 
Tracing over only long wavelength modes, the number of such modes
is proportional to the spatial volume. So it is only natural 
that $N$ should 
be proportional to the $9$-volume $V_{9}$. Writing $V_{9}$ as 
$V_{6}\bar{a}^{3}$, we have argued in Sec. 2 that 
$\bar{a}^{3} = a a^{\prime 2}$ for consistent decoherence to take place.
Any other choice would lead to either zero or no decoherence
Here we are staying in the Euclidean region and starting with a finite 
$a^{\prime}$ and $a$; we then take the limit where $a^{\prime} \to 0$. 
Decoherence is seen if we choose the cut-off scheme 
$N \propto V_{6}a a^{\prime 2}$. 
Strictly speaking, the whole calculation in quantum gravity is 
ill-defined at $a^{\prime} = 0$ where classical space-time does not exist. 
Another way to see the difficulty is that, close to $a^{\prime} = 0$, the 
separation between lower and higher multipoles becomes less and 
less accurate. The entire perturbative calculation done 
in Ref.\cite{Halliwell:1985eu} breaks down at $a^{\prime} = 0$. 
Since classical spacetime are merely derived (i.e., not fundamental)
quantities in string theory, the $a^{\prime}=0$ limit should be 
meaningful and hopefully calculable.
There have been discussions about the calculation of the wavefunction 
of the universe in a string theory context in\cite{ Horowitz:2003he}.

 Perhaps we can convince the reader that the decoherence term 
$\cal{D}$$ \propto V_{9}$ by considering the case where the universe 
starts off at some finite $a^{\prime}$ with energy less than the height 
of the barrier as described by a nonvanishing $\epsilon$ parameter 
in Eq.(\ref{two}). After tunnelling, it becomes a deSitter state on 
the other side of the barrier in Figure 6. The calculation of the 
decoherence term can now be done with tensor modes and the final 
result should have $\cal{D}$$ \propto V_{9}$. 
We expect the qualitative result $\cal{D}$ $ \propto V_{9}$ to be 
true for this analysis. 
One can then take the limit $\epsilon \to 0$ and obtain the result 
for the tunneling from nothing. 
A macroscopic universe would not tunnel because of decoherence. 



\section{$S^{10}$ with Dilaton}

\begin{figure}
\begin{center}
\epsfig{file=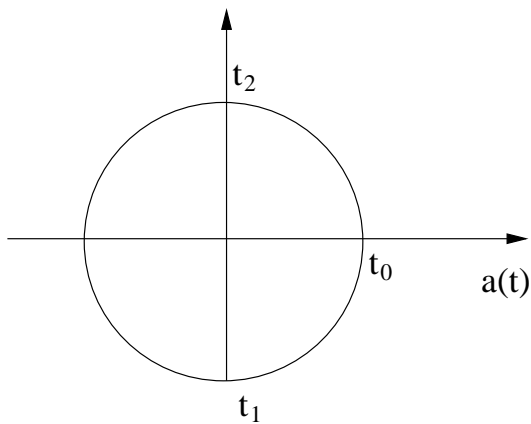, width=7cm}
\caption{$S^4$ instanton  }
\label{S10wD}
\end{center}
\end{figure}

In this section we investigate the impact of dilaton on the action for 
$S^{10}.$ The action with dilaton included in string frame is
\ba
S= \frac{1}{16 \pi G} \int d^{10}x \sqrt{|g|} e^{-2\phi} \left(R - 
2\Lambda \, e^{\phi} +4(\nabla \phi)^2 \right)
\ea
After the Weyl transformation 
$g_{mn}\rightarrow e^{\frac{\phi}{2}} g_{mn} $, 
the equations in Einstein frame are
\ba
\label{00eq}
36 \left( \frac{\dot{a}^{2}}{a^{2}} - \frac{1}{a^{2}}\right)
 &=& -\Lambda e^{\frac{3\phi}{2}}+\frac{1}{4}{\dot {\phi}}^2
\\ 
\label{iieq}
8\frac{\ddot{a}}{a}+
28 \left( \frac{\dot{a}^{2}}{a^{2}} - \frac{1}{a^{2}}\right) &=&
-\Lambda e^{\frac{3\phi}{2}}-\frac{1}{4}{\dot {\phi}}^2 \\
\label{dil}
\ddot {\phi}+9\frac{\dot{a}}{a}\dot{\phi}&=&3\Lambda e^{\frac{3\phi}{2}}
\ea
We are interested in the instanton solution with the following properties:
\ba
a(t_1)=a(t_2)=0 \\
a(t)\le a(t_0)\nonumber\\
\dot{a}|_{t_0}=\dot{\phi}|_{t_0}=0 \nonumber
\ea
The first two conditions correspond to having a closed instanton with 
spherical topology. The third condition comes from the analyticity 
and real-valued properties of 
the solution in transition from Euclidean space-time to the Lorentzian 
space-time (see Figure \ref{S10wD}). 
We show that, with the dilaton included, there are apparent singularities 
at $t=t_1$ and $t=t_2$. In other words, $t_2(t_1)$ corresponds to 
big-bang (big-crunch).

To begin the proof, suppose the solution is non-singular. 
The point of special  
care are $t_1$ and $t_2$, where $a(t)=0$. We focus our investigations near 
$t=t_2$. The same discussion applies equally to the $t_1$ case.
To have a regular solution, all 
scalars like $R$, $R_{mn}R^{mn}$, $R_{mnpq}R^{mnpq},...$ constructed from 
the metric components and their derivatives must be regular. 
In particular, we have
\ba
R_{mnpq}R^{mnpq}=36\left[(\frac{\ddot{a}}{a})^2+
4\left((\frac{\dot{a}}{a})^2-\frac{1}{a^2} \right )^2 \right]
\ea
Clearly, $R_{mnpq}R^{mnpq}$ is regular if both $(\frac{\ddot{a}}{a})$ and 
$(\frac{\dot{a}}{a})^2-\frac{1}{a^2}$ are regular. Since $a(t)$ vanishes at
$t_2$ by construction, $\frac{\dot{a}}{a}$ will diverge such that 
$(\frac{\dot{a}}{a})^2-\frac{1}{a^2}$ is finite. Suppose
\ba
\label{asing}
a(t) \sim \alpha[(t-t_2)^p +\beta(t-t_2)^q]~~~~~, ~~~~~0<p<q \ .
\ea
We find
\ba
\frac{\dot{a}}{a} &\sim& p(t-t_2)^{-1}+\beta (q-p)(t-t_2)^{q-p-1}\nonumber\\
\frac{\ddot{a}}{a}& \sim& p(p-1)(t-t_2)^{-2}+
\beta\left[q(q-1)-p(p-1)\right](t-t_2)^{q-p-2}
\ea
To have $(\frac{\dot{a}}{a})^2-\frac{1}{a^2}$ and 
$\frac{\ddot{a}}{a}$ regular, we need ${\alpha}=-1$ and $q-p \ge 2$, 
respectively. This implies that $q\ge 3$. So 
\ba
\label{aaprox}
a(t)\sim (t_2-t)+\beta(t-t_2)^q  ~~~~~~~~, ~~~~~q \ge 3.
\ea
On the other hand
\ba
\label{dotphi}
{\dot{\phi}}^2=16\left((\frac{\dot{a}}{a})^2-\frac{1}{a^2} 
-\frac{\ddot{a}}{a} \right)
\ea
Near $t_2$, by using Eq(\ref{aaprox}), we find
\ba
{\dot{\phi}}^2 \sim 8\beta q(3-q)(t-t_2)^{q-3} \ .
\ea 
Previously we concluded that $q \ge 3$, which indicates that 
$\dot \phi \rightarrow 0$ with some positive power.

Now consider the scalar field equation. It is more instructive 
to write Eq(\ref{dil}) in the following form
\ba
\frac{d}{dt}{(a^9\phi)}=3\Lambda a^9 e^{\frac{3\phi}{2}}
\ea
Define $F(t)\equiv a^9 \dot{\phi}$. At $t=0$, we demand $\dot{\phi}=0$, so 
$F(0)=0$. Near $t=t_2$, the regularity assumption of the solution requires 
that $a(t)\sim (t_2-t), \dot{\phi} \rightarrow 0$, which implies that 
$F(t_2)=0$ as well. With the conditions $F(0)=F(t_2)=0$, the mean value 
theorem of analysis implies that $\dot {F}$ vanishes 
at some point in the interval 
$[0,t_2]$. But $\dot {F}=3\Lambda a^9 e^{\frac{3\phi}{2}}$ is always 
positive definite. 
We conclude that regularity at $t=t_2$ is a false assumption and the system 
develops a singularity at $t=t_2$. It is easy to see that the only 
property 
of the dilaton exponential potential which was used to verify the 
existence of the 
singularity is the fact that $\frac{\partial V}{\partial{\phi}}>0$. 
This means 
that the singularity is a generic properties of any scalar field with 
monotonic potential.

In order to evaluate the action we need to find the form of the 
singularity. Using Eq.(\ref{asing}), from Eq.(\ref{dotphi}) we find
\ba
{\dot{\phi}}^2 \sim 16\left(\frac{p}{(t-t_2)^2} -
\frac{1}{\alpha^2(t-t_2)^{2p}}\right)
\ea
To have ${\dot{\phi}}^2>0$,  one requires $p\le 1$, which  
implies that $\dot{\phi}\sim \frac{1}{t-t_2}$ up to a coefficient.
The possibility of $p=1$ is excluded. The 
previous discussion indicates that $F(t)\equiv a^9\dot{\phi}$ 
should not vanish 
at $t=t_2$, which is not possible for $a(t)\sim (t-t_2)$ and 
$\dot{\phi} \sim \frac{1}{t-t_2}$. This will fix the coefficient of 
$\dot{\phi}$ and we find $\dot{\phi}=\pm \frac{4\sqrt{p}}{(t-t_2)}$ and 
$\phi \sim \pm 4 \sqrt{p} \, \ln|t-t_2|$. Using the scalar field 
equation, one can easily see that the positive branch solution for $\phi$ 
is not allowed and matching the most singular power of $\frac{1}{(t-t_2)}$ 
will fix $p=\frac{1}{9}$. In summary, we find
\ba
a(t)\sim (t_2-t)^{\frac {1}{9}} ~~~~~~, ~~~~~~ 
\phi(t) \sim \frac{4}{3}ln|t-t_2| \ .
\ea
To evaluate the action, from Eqs(\ref{00eq}) and (\ref{iieq}), $\dot{\phi}$ 
and $\Lambda e^{\frac{3\phi}{2}}$ can be expressed in terms of $a(t)$ 
and its derivatives. We find
\ba
S=\frac{1}{16\pi G}\int d^{10}x \, a^9 \left[2 (\frac{\ddot{a}}{a})
+16\left((\frac{\dot{a}}{a})^2-\frac{1}{a^2} \right)
 \right]
\ea
Near $t=t_2$, the singular terms in the bracket cancel each other and the 
integrand behaves like $(t_2-t)^{\frac{7}{9}}$ and the action is finite.

Although the analysis are carried out only for $S^{10}$, one can expect the
same qualitative result apply for other instantons considered in section 3 and
the dilaton inclusion will not change the result.

\section{$S^4 \times S^6$ and Other Instantons}

We start with the 10-dimensional Euclidean metric ansatz
for $S^{1+n_{1}} \times S^{n_{2}}$ ($n_1 + n_2 +1=10$) with the
metric (\ref{Sn1Sn2}).
The Einstein equation $G_{\mu \nu} = 8 \pi G_{10} T^{vac}_{\mu \nu}=
-(\Lambda_a g^{(1 + n_1)}_{\mu \nu},
\Lambda_b g^{(n_2)}_{\mu \nu})$ 
becomes (the time-time component, the space-space component in 
$S^{1+n_1}$ and the space-space component in the $S^{n_2}$ respectively)
\baray
-\frac{n_{1}(n_{1}-1)}{2}\left(\frac{\dot{a}^{2}}{a^{2}} - \frac{1}{a^{2}}
\right) - \frac{n_{2}(n_{2}-1)}{2}\left(\frac{\dot{b}^{2}}{b^{2}} 
- \frac{1}{b^{2}}\right) - n_{1}n_{2}\frac{\dot{a}}{a}\frac{\dot{b}}{b} &= 
\Lambda_a \\ \nonumber
-(n_{1}-1)\frac{\ddot{a}}{a}- n_{2}\frac{\ddot{b}}{b}- 
\frac{(n_{1}-1)(n_{1}-2)}{2}\left(\frac{\dot{a}^{2}}{a^{2}} - 
\frac{1}{a^{2}}\right)  & \\ \nonumber 
-\frac{n_{2}(n_{2}-1)}{2}\left(\frac{\dot{b}^{2}}{b^{2}} - 
\frac{1}{b^{2}}\right)  
-(n_{1}-1)n_{2}\frac{\dot{a}}{a}\frac{\dot{b}}{b} &= \Lambda_a \\ \nonumber
-n_{1}\frac{\ddot{a}}{a}- (n_{2}-1)\frac{\ddot{b}}{b}- 
\frac{n_{1}(n_{1}-1)}{2}\left(\frac{\dot{a}^{2}}{a^{2}} -
\frac{1}{a^{2}}\right) & \\ \nonumber 
-\frac{(n_{2}-1)(n_{2}-2)}{2}\left(\frac{\dot{b}^{2}}{b^{2}} - 
\frac{1}{b^{2}}\right) 
- n_{1}(n_{2}-1)\frac{\dot{a}}{a}\frac{\dot{b}}{b}
&= \Lambda_b 
\earay
where the dot is with respect to Euclidean time $\tau$.
(Different $\Lambda$s may be realized by having some $D9-\bar D 9$
branes plus, for example, $D(1+n_1)-\bar D (1+n_1)$ pairs of branes
behaving like dust in the extra $n_2$ dimensions.) 
The instanton solution corresponds to a $S^{1+n_{1}}$ bounce 
with a static $S^{n_{2}}$ (constant $b(\tau)= b$):
\baray
\label{bounce}
-{\dot a}^2 +1 =H^2 a^2 \\ \nonumber
a(\tau)~=~ H^{-1}\cos(H\tau) \\ \nonumber
H^2 = \frac{2(n_1-1)\Lambda_a -n_2(n_1-1)((\Lambda_a - \Lambda_b)}
{n_1(n_1-1)(n_1 +n_2 -1)}\\ \nonumber
b^2 = \frac{(n_2-1)(n_1 + n_2 -1)}
{(\Lambda_a + \Lambda_b) +n_1(\Lambda_a - \Lambda_b)}
\earay
Here $S^{n_{2}}$ is cosmologically stabilized, with radius $b$.
For $\Lambda_a = \Lambda_b= \Lambda$, the above result reduces to 
\baray
\label{s4s6}
b^2~=~ \frac{(n_{2}-1)(n_{1}+n_{2}-1)}{2\Lambda} \\ \nonumber
H^{2} ~=~ \frac{2\Lambda}{n_{1}(n_{1}+n_{2}-1)}
\earay
For the case $n_2=0$, the last component of the above Einstein equation 
is absent and we have $H^{2}= 2\Lambda/{n(n-1)}$.
For $S^{10}$, this becomes $H^{2}= \Lambda/36$. 

For the case of $S^{4} \times S^{6}$ with equal $\Lambda$, 
$n_{1}=3$ and $n_{2}=6$, and 
we get $ b~=~ \sqrt{20/\Lambda}$  and $H^{2} ~=~ \Lambda/12$. 
By Wick rotating the time axis to Minkowski metric: 
$\tau \rightarrow it$, we have
\baray
a(t)~=~ H^{-1}\cosh(Ht)
\earay
this is just the usual $1+3$ dimensional deSitter space-time with a 
static $S^6$ making up the extra dimensions. It describes a universe 
that is contracting at $t < 0$, reaches its minimum size 
($a_{minimum}=1/H$) at $t=0$ and then expands for $t >0$.
For large $t$, the expansion is exponentially, corresponding
to an inflationary universe. Since the cosmological constant is 
due to brane-anti-brane pairs, this universe is always classically 
unstable. The presence of tachyonic modes will lead to their eventual
annihilations, which will end inflation.
On the other hand, the above Euclidean solution describes a 
compact space; it is defined only for $|\tau| \le \pi/2H$.
Together, they describe the quantum tunneling process (See Figures 
1 and 2). Starting from ``nothing'' (i.e., no space and no time), 
the instanton solution bounces at the classical turning point $a=1/H$
(where $\dot a=0$), and continues at Minkowski time $t$. 

The Euclidean action $S_E$ and the spatial volume are easy to determine.
As an example, consider the case of $S^{4} \times S^{6}$. 
The geometrical factor $\int \sqrt{g}d^{10}x$ factorizes to 
$A_{4} \times A_{6}$. The radius of $S^{4}$ is just the Hubble radius, 
$H^{-1}$. The radius of $S^{6}$ is $ b~=~ \sqrt{20/\Lambda}$, so
 $A_{4} = \frac{8}{3}\pi^{2}H^{-4}$ and 
$A_{6} = \frac{16}{15}\pi^{3}b^{6}$. So 
\baray
 S_{E}= -\frac{1}{16\pi G_{10}} \times \frac{5\Lambda}{2} \times \int \sqrt{g}d^{10}x = \frac{1}{16\pi G_{10}} \times  \frac{5\Lambda}{2} \times A_{4} \times A_{6} \\ \nonumber = \frac{1}{16\pi G_{10}} \times  \frac{5\Lambda}{2} \times \frac{8}{3}\pi^{2}(H^{-1})^{4} \times \frac{16}{15}\pi^{3}b^{6} = 1.57187 \times 10^{10}
\earay

The generalization to  
$S^{n_{1}+1} \times S^{n_{2}} \times ... \times S^{n_{k}}$ instantons
is completely straightforward. One finds:
\baray
H^{2}~=~ \frac{2\Lambda}{n_{1}(D-2)}\\ \nonumber
a_{1}(t)~=~ H^{-1}\cos(Ht) \\ \nonumber
a_{l}~=~ \sqrt{\frac{(n_{l}-1)(D-2)}{2\Lambda}}
\earay
where $l= 2,3,....k$ and $D=\sum_{i=1}^{k}n_{i}+1 $. The cosine 
dependence of $S^{n_{1}+1}$ becomes a $\cosh$
dependence in Lorentzian space-time and the solution describes, after tunneling, inflating $n_{1}+1$ dimensional space-time with $\sum_{i=2}^{k}n_{i}$ extra dimensions that are static and have the geometry of $ S^{n_{2}} \times ... \times S^{n_{k}}$. 

One may also consider instantons with torus fibered over sphere 
: $S^{n_{1}+1} \times T^{1}S^{n_{2}} \times T^{m}$.
For example,  consider $T^{1}$ fibered over $S^{n_{2}}$ as implied 
by the notation $... \times T^{1}S^{n_{2}} \times... $. 
The metric is given by:
\baray
ds^{2} =&  d\tau^{2} + a(\tau)^{2}\left( \frac{dr^{2}}{1-r^{2}} +  
r^{2}d\Omega^{2}_{(n_1-1)}\right) \\ \nonumber
&+ b^{2} \left( \frac{d\rho^{2}}
{1-\rho^{2}} + \rho^{2}d\Omega^{2}_{(n_2-1)}\right) + c(\theta)^{2}d\alpha^{2} + d^{2}\sum_{i=1}^{m}d\beta_{i}^{2}
\earay
where $\alpha$ and $\beta_{i}$ are the scale factors of the torii. 
$\theta$ is the azimuthal coordinate of $S^{n_{2}}$ and the $\theta$ 
dependence of the torus scale factor $c(\theta)$ expresses the 
fibration of the torus $T^{1}$ on $S^{n_{2}}$. $c(\theta)$ and $d$
are taken to be time independent as we are looking for solutions
where the extra dimensions are time independent.
The Euclidean Einstein equations for this metric are:
\baray
-\frac{1}{2}n_{1}(n_{1}-1)\left(\frac{\dot{a}^{2}}{a^{2}}-\frac{1}{a^{2}}\right)-\frac{1}{2}n_{2}(n_{2}-1)\frac{1}{b^{2}}- \frac{1}{b^{2}}\left(\cot{\theta}\frac{c^{\prime}(\theta)}{c(\theta)}+\frac{c^{\prime \prime}(\theta)}{c(\theta)} \right) = \Lambda \nonumber \\
-\frac{\ddot{a}}{a}+(n_{1}-1)\left(\frac{\dot{a}^{2}}{a^{2}}-\frac{1}{a^{2}}\right)=0 \nonumber \\
-n_{1}\frac{\ddot{a}}{a}-\frac{1}{2}n_{1}(n_{1}-1)\left(\frac{\dot{a}^{2}}{a^{2}}-\frac{1}{a^{2}}\right)-\frac{1}{2}(n_{2}-1)(n_{2}-2)\frac{1}{b^{2}}- \frac{1}{b^{2}}\left(\cot{\theta}\frac{c^{\prime}(\theta)}{c(\theta)} \right) = \Lambda \nonumber \\
-n_{1}\frac{\ddot{a}}{a}-\frac{1}{2}n_{1}(n_{1}-1)\left(\frac{\dot{a}^{2}}{a^{2}}-\frac{1}{a^{2}}\right)-\frac{1}{2}(n_{2}-1)(n_{2}-2)\frac{1}{b^{2}}- \frac{1}{b^{2}}\left( +\frac{c^{\prime \prime}(\theta)}{c(\theta)}  \right) = \Lambda \nonumber \\
-n_{1}\frac{\ddot{a}}{a}-\frac{1}{2}n_{1}(n_{1}-1)\left(\frac{\dot{a}^{2}}{a^{2}}-\frac{1}{a^{2}}\right) -\frac{1}{2}n_{2}(n_{2}-1)\frac{1}{b^{2}} = \Lambda \nonumber \\
\frac{\dot{b}}{b}\frac{c^{'}}{c}- \frac{1}{c}\frac{\partial^{2}c}{\partial{\theta}\partial{t}} = 0 
\earay
 
The solution of Euclidean Einstein equations for this metric is :
\baray
\label{fiber}
a(t)~=~ H^{-1}\cos(Ht), \\ \nonumber
\frac{1}{b^{2}}~=~ \frac{2\Lambda}{[2(n_{1}+1)+n_{2}(n_{2}-1)]}, \\ \nonumber
c(\theta) = \cos(\theta), \\ \nonumber
H^{2}~=~ \frac{2}{n_{1}b^{2}}
\earay 

 The static scale factor 
$d$ of the torus $T^{m}$ is not fixed by these equations.

One may consider $S^{1+n_{1}}\times T^{1}S^{2} \times T^{m}$ type instantons in $10-D$ $(1+n_{1}+1+2+m = 10)$ spacetime and see how the action varies. Using (\ref{fiber}) one can find the action for such an instanton. The action for an instanton  $S^{1+n_{1}}\times T^{1}S^{n_{2}} \times T^{m}$ in D-spacetime dimensions is given by:
\baray
\lefteqn{S_{E} = -\frac{1}{16 \pi G^{D}}\times \frac{4\Lambda}{D-2} \times Volume} \nonumber \\
&&  =   -\frac{4\Lambda}{16 \pi G^{D}(D-2)}
 \frac{2 \pi^{(2+n_{1})/2}(H^{-1})^{1+n_{1}}}{\Gamma(\frac{2+n_{1}}{2}) }
\frac{2 \pi^{(2+n_{2})/2}(b)^{n_{2}}}{\Gamma(\frac{2+n_{2}}{2}) } 
(\frac{2\pi A}{M_{s}})^{1+m}
\earay
where $A$ is the size of the torii in string units. Using (\ref{fiber}), (\ref{newton}), (\ref{vac}), for the case $D=10$ and $n_{2}=2$ one obtains: 
\baray
S_{E} = -\left( 16\pi A^{7}\right) \frac{1}{n_{1}A^{n_{1}} (n_{1}/2)!} 
\left( \frac{\pi}{2}n_{1}(2+n_{1}) \right)^{(3+n_{1})/2}
\earay

\section{Locating the Preferred Inflationary State}

Consider the KKLMMT scenario. The K\"{a}hler potential is
\ba
{\cal{K}}=-3 \ln [-i(\rho-\rb)]- \ln[-i(\tau-\bar{\tau})]-
\ln\left(-i\int_{\cal{M}}\Omega \wedge \bar{\Omega} \right).
\ea
The supersymmetric minimum is given by $D_zW=D_{\tau}W=D_{\rho}W=0$
(where $D_iW=\partial_iW + \partial_i{\cal{K}} W$), which 
lead to, respectively, 
\ba
\label{susyeq1}
\frac{M}{2\pi i}(1+ \ln z)-(K+K'\frac{\partial z'}{\partial z} )\tau
&=&0 \\
\label{susyeq2}
M{\cal{G}}-(Kz+K'z')\bar{\tau}&=&-Ae^{ia\rho} \\
\label{susyeq3}
M{\cal{G}}-(Kz+K' z')\tau &=&\left(\frac{i}{3}a(\rho-\bar{\rho})-1\right) 
Ae^{ia\rho}
\ea
Here we present solutions of Eqs(\ref{susyeq1}), (\ref{susyeq2})
and (\ref{susyeq3}).
To simplify the analysis we take $\tau=ie^{-\phi}=i/g_s$, 
$\rho_{min}=i\sigma$ while $z$ and $z'$ are taken to be real. 
Also we assume 
$z'(z)\sim {\cal{O}}(1) $ and $\partial z'(z)/ \partial z \sim 0$ for small 
$z$. In the limit of physical interest where $K/g_s$ is 
large, we find from Eq.(\ref{susyeq1})  that
\ba
\label{zexpression}
z= \exp \left( -\frac{2\pi K}{Mg_s} \right) \ . 
\ea
In this limit, from the sum and difference 
of the remaining two equations, respectively, we find
\ba
\label{sigma}
& a\sigma = -3-LW(-1,-m) \\ 
\frac{1}{g_s} =& -\frac{i}{3K'z'(0)} A \, a\sigma \exp(-a \sigma)
=\frac{|{\cal{G}}(0)| M}{ K'z'(0)}\, \beta(m)
\label{g_s}
\ea 
where we have introduced $m \equiv 3e^{-3}M|{\cal{G}}(0)A^{-1}|$ and 
$\beta(m)\equiv \frac{3+LW(-1,-m)}{LW(-1,-m)}$. To have a solution, 
we need $m < 3e^{-3} \sim 0.15$. 
The function $ y(x)= -LW(-1,-x)$ is the LambertW functions, which is
the solution of the equation 
$ye^{-y}=x, x<e^{-1}$, for  $y>1$. 
Using Eq.(\ref{g_s}) in $z$ expression, Eq.(\ref{zexpression}), we find
\ba
\label{z}
z=e^{-2\pi |{\cal{G}}(0)| \beta(m) K/z'(0)K'} \ .
\ea
We are interested in maximizing ${\F} \equiv -S_E -{\hat c} M_s^9 V_9$ 
in terms of fluxes, $K, M$ and $K'$.
Using the solutions found for $z, \rho(\sigma)$ and $g_s$ and using 
Eqs (\ref{rhoG2}) and (\ref{v9expression}) for $S_E$ and $c M_s^9 V_9$ ,
respectively, we find
\ba
\label{finalF}
{\F} = \bar A f_1(m)e^{\frac{8\pi |{\cal{G}}(0)| K\beta(m)}{3K'}}-
{\hat c} \bar B f_2(m)e^{\frac{4\pi |{\cal{G}}(0)|  K\beta(m)}{K'}}
\ea
where 
\ba
\bar A &\equiv&\frac{12 }{(2\pi)^9}
\left(\frac{e^3}{3 z'(0)}\right)^3 |A|^3 a^{-3}
\nonumber\\
\bar B &\equiv& \frac{2 \sqrt3}{(2\pi)^5}
\left(\frac{e^3}{3 z'(0)}
\right)^{\frac{3}{2}}|A|^{\frac{3}{2}} a^{-\frac{15}{4}}
 \nonumber\\
f_1(m)&\equiv& m^3 \frac{\left(3+LW(-1,-m)\right)^6}{|LW(-1,-m)|^3}\nonumber\\
f_2(m)&\equiv& m^{\frac{3}{2}} \frac{\left|3+LW(-1,-m)\right|^{\frac{21}{4}}}
{|LW(-1,-m)|^{\frac{3}{2}}}
\ea
From the combination of 
$\frac{\partial }{\partial K}{\F}=0$ and
$\frac{\partial }{\partial M}{\F}=0$, we obtain
\ba
\label{quantized}
3f_1'f_2-2 f_1f_2'&=&0 \\\nonumber
\label{expp}
e^{\frac{pK\beta(m)}{2K'}}&=&\frac{2\bar A}{3c\bar B}\frac{f_1(m)}{f_2(m)}
\ea
The root of the first equation, independent of $a$ and $|A|$ is 
$m_{max}\sim 0.06$.
It turns out that the condition $\frac{\partial }{\partial K'}{\F}=0$ 
is not quite compatible with $\frac{\partial }{\partial K}{\F}=0$. 
On the other hand, $K'$ appears only in the denominator of the exponential 
terms in ${\F}$. Since the fluxes are quantized we can choose the 
smallest possible integer value for $K'$ to maximize ${\F}$; that is, 
$K'=1$.

From Eqs (\ref{quantized}) and (\ref{expp}), respectively, we find
\ba
\label{Mexact}
M&=&\frac{m \, e^3}{3}|\frac{A}{{\cal{G}}(0)}|\\\nonumber
\label{Kexact}
K&=&\frac{3}{4\pi{\cal{G}}(0)\beta(m)}\ln
\left( \frac{4\sqrt3 \, a^{\frac{3}{4}}}{(2\pi)^4}
\frac{f_1(m)}{f_2(m)} \,
|\frac{e^3A}{3z'(0)}|^{\frac{3}{2}}\frac{1}{\hat{c}}
\right) \ .
\ea
Correspondingly, the warp factor is given by
\ba
(\frac{r_0}{R})\sim z^{1/3}=
\left( \frac{4\sqrt3 \, a^{\frac{3}{4}}}{(2\pi)^4}
\frac{f_1(m)}{f_2(m)} \,
|\frac{e^3A}{3z'(0)}|^{\frac{3}{2}}\frac{1}{\hat{c}} \right)^{-1/2}
\ea
From 
Eqs (\ref{Mpl}), (\ref{g_sapprox}) and (\ref{Mvalue}), we find
\ba
(\frac{M_{Pl}}{M_s})^2=\frac{2\, e^6 m^2\beta(m)^2}{9\, (2\pi)^7z'(0)}\, 
|3+LW(-1,-m)|^{\frac{3}{2}}\, a^{-\frac{3}{2}}|A|^2 \ .
\ea 
To get a better understanding of the results, we will use the approximate
values for $\beta(m)$, $f_1(m)$ and $f_2(m)$ at the maximum point 
$m_{max}\sim 0.06$:
\ba
\beta(m)\sim 0.3~~~~~,~~~~~f_1(m)\sim 10^{-5}~~~~~,~~~~~
f_2(m)\sim 5\times 10^{-3} \ .
\ea
Also the geometric quantities $|{\cal{G}}(0)|$ and $z'(0)$ are basically 
calculable. We use the GKP approximation and set 
$|{\cal{G}}(0)| \sim z'(0) \sim 1$, so
 the expressions for $M$, $K$, the warp factor and 
$M_{Pl}/M_s$, respectively, simplify to
\ba
\label{Mvalue}
M &\sim&  0.4 \times |A| \\\nonumber
K &\sim& 0.8 \times \ln\left(10^{-4} \, a^{3/4}\, |A|^{3/2}\,
{\hat c}^{-1} \right)\\\nonumber
\label{warpapp}
(\frac{r_0}{R})&\sim&115 \times a^{-3/8}\, |A|^{-3/4}\, 
{\hat c}^{1/2}\\\nonumber
\frac{M_{Pl}}{M_s}&\sim& 3\times 10^{-4}\,  a^{-3/4}\, |A|
\ea
The value of function ${\F}$ at the maximum is
\ba
\label{Fmax}
{\F}_{max} \sim 2\times 10^{-18}\, a^{-3/2}\, |A|^6\, {\hat c}^{-2} \ .
\ea
Finally, the COBE normalization (that is, the density perturbation
$\delta_H \simeq 10^{-5}$) is given in Eq.(C.8) of KKLMMT as
\ba
\label{COBE}
\delta_H=C_1 N_e^{5/6}\, (\frac{T_3}{M_p})^{1/3}\, (\frac{r_0}{R})^{4/3} \ ,
\ea
where $N_e$ is the number of e-folding and $C_1\sim 0.4$ in their model.
Using Eqs (\ref{D3tension}), (\ref{Mpl}) and (\ref{warpapp}) 
and taking $N_e \sim 60$, we find 
\ba
\label{COBE2}
\delta_H\sim 2 \times 10^7 \,  a^{1/2}\, |A|^{-2}\, {\hat c}^{2/3} \ .
\ea
There is an interesting relation between ${\F}$ and $\delta_H$. 
One can easily show from Eqs (\ref{Fmax}) and (\ref{COBE2}) that 
\ba
{\F}_{max} \sim 1.6 \times 10^4 \, \delta_H^{-3} \ .
\ea 
For $\delta_H\sim  10^{-5}$, we find
\ba
{\F}\sim 10^{18} \ .
\ea
This value for ${\F}$ is many order of magnitudes bigger than the value 
of ${\F}$ for $S^{10}$.

To check the validity of the low energy supergravity limit, 
we must verify that $g_s<1$ and $r>>1$(in the unit of $M_s$). 
From Eqs (\ref{g_s}) and (\ref{Mvalue}) we find
\ba
\label{g_sapprox}
\frac{1}{g_s}=\frac{s M}{2\pi K'z'(0)}\, \beta(m) \,  \sim 0.1\times |A| 
\ea 
To satisfy the bound $g_s<1$, we take $|A|>10$. From Eq.(\ref{sigma}) 
we have
\ba
|\sigma|= r^4=-\frac{1}{a}\left(3+LW(-1,-m)\right) \, \sim \frac{1.2}{a}
\ea
where the approximation $-3-LW(-1,-m)\sim 1.2$ was used.
We see that to have $r>>1$, we need $a<<1$. For example the value chosen
in KKLT for $a=0.1$, results in $r\sim 2$. \\

It is easy to generalize the above inflationary scenario to 
$\cal{N}$ pairs of $D3-\bar{D}$3 branes. 
We also add a very small $\Lambda_{\text{KKLT}}$
like today's cosmological constant as
is obtained in KKLT:
\ba
\Lambda \rightarrow \hat{\Lambda}= 8\pi G {\cal{N}}\,  T_3\,  
(\frac{r_0}{R})^4+\Lambda_{\text{KKLT}} \ .
\ea
For ${\cal{N}} \neq 0$, the first term clearly dominates and
the second term may be ignored. This 
corresponds to $\bar A \rightarrow {\cal{N}}^{-1}\bar A$ and
$\bar B \rightarrow {\cal{N}}^{-3/2} \bar {B} $ in Eq.(\ref{finalF}). 
The extremum can be obtained exactly as before. $M$ is given as in 
Eq.(\ref{Mexact}) while $K$ is given by
\ba
K \rightarrow K+ \frac{3}{8\pi{\cal{G}}(0)\beta(m)}\ln{{\cal{N}}} \ .
\ea
So we see that the flux $K$ is not very sensitive to the value of 
${\cal{N}}$.

To compare the inflationary states with the KKLT vacua, we 
calculate ${\F}$ for ${\cal{N}}= 0$:
\ba
\label{FKKLT}
{\F}_{KKLT} =\frac{3\pi}{2G_N \Lambda_{KKLT}}
\left[1- \frac{(2\pi)^6}{2\sqrt3}\, \hat{c}\, g_s^2 
\frac{M_s}{\sqrt{\Lambda_{KKLT}}}\right]
\ea 
To find an estimate for ${\F}_{KKLT} $, let take 
$g_s \sim 10^{-1}$, $\hat c \sim 10^{-10}$ and 
$\Lambda_{KKLT}/M_{Pl}^2\sim 10^{-120}$ corresponding to today's 
cosmological constant. One can easily verify that the $\cal D$ term is 
exponentially larger than $-S_E$, and
\ba
{\F}_{KKLT}\sim -10^{172}\, (\frac{M_s}{M_{Pl}}) \ .
\ea
This means that the tunneling to a small $\Lambda$ universe is 
severely suppressed, in contrast to that suggested by 
the Hartle-Hawking scenario.


\begin{thebibliography}{10}

\bibitem{Giddings:2001yu}
S.~B. Giddings, S.~Kachru, and J.~Polchinski, {\it Hierarchies from fluxes in
  string compactifications},  {\em Phys. Rev.} {\bf D66} (2002) 106006,
  [\href{http://arXiv.org/abs/hep-th/0105097}{{\tt hep-th/0105097}}].

\bibitem{Kachru:2003aw}
S.~Kachru, R.~Kallosh, A.~Linde, and S.~P. Trivedi, {\it De sitter vacua in
  string theory},  {\em Phys. Rev.} {\bf D68} (2003) 046005,
  [\href{http://arXiv.org/abs/hep-th/0301240}{{\tt hep-th/0301240}}].

\bibitem{Saltman:2004sn}
A.~Saltman and E.~Silverstein, {\it The scaling of the no-scale potential and
  de sitter model building},  \href{http://arXiv.org/abs/hep-th/0402135}{{\tt
  hep-th/0402135}}.

\bibitem{Denef:2004dm}
F.~Denef, M.~R. Douglas, and B.~Florea, {\it Building a better racetrack},
  \href{http://arXiv.org/abs/hep-th/0404257}{{\tt hep-th/0404257}}.

\bibitem{Denef:2004ze}
F.~Denef and M.~R. Douglas, {\it Distributions of flux},
  \href{http://arXiv.org/abs/hep-th/0404116}{{\tt hep-th/0404116}}.

\bibitem{Douglas:2003um}
M.~R. Douglas, {\it The statistics of string / m theory vacua},  {\em JHEP}
  {\bf 05} (2003) 046, [\href{http://arXiv.org/abs/hep-th/0303194}{{\tt
  hep-th/0303194}}].

\bibitem{Ashok:2003gk}
S.~Ashok and M.~R. Douglas, {\it Counting flux vacua},  {\em JHEP} {\bf 01}
  (2004) 060, [\href{http://arXiv.org/abs/hep-th/0307049}{{\tt
  hep-th/0307049}}].

\bibitem{Douglas:2004kp}
M.~R. Douglas, {\it Statistics of string vacua},
  \href{http://arXiv.org/abs/hep-ph/0401004}{{\tt hep-ph/0401004}}.

\bibitem{Weinberg:2000qm}
S.~Weinberg, {\it A priori probability distribution of the cosmological
  constant},  {\em Phys. Rev.} {\bf D61} (2000) 103505,
  [\href{http://arXiv.org/abs/astro-ph/0002387}{{\tt astro-ph/0002387}}].

\bibitem{Susskind:2003kw}
L.~Susskind, {\it The anthropic landscape of string theory},
  \href{http://arXiv.org/abs/hep-th/0302219}{{\tt hep-th/0302219}}.

\bibitem{Susskind:2004uv}
L.~Susskind, {\it Supersymmetry breaking in the anthropic landscape},
  \href{http://arXiv.org/abs/hep-th/0405189}{{\tt hep-th/0405189}}.

\bibitem{Banks:2003es}
T.~Banks, M.~Dine, and E.~Gorbatov, {\it Is there a string theory landscape?},
  \href{http://arXiv.org/abs/hep-th/0309170}{{\tt hep-th/0309170}}.

\bibitem{Peiris:2003ff}
H.~V. Peiris {\em et.~al.}, {\it First year wilkinson microwave anisotropy
  probe (wmap) observations: Implications for inflation},  {\em Astrophys. J.
  Suppl.} {\bf 148} (2003) 213,
  [\href{http://arXiv.org/abs/astro-ph/0302225}{{\tt astro-ph/0302225}}].

\bibitem{Guth:1981zm}
A.~H. Guth, {\it The inflationary universe: A possible solution to the horizon
  and flatness problems},  {\em Phys. Rev.} {\bf D23} (1981) 347.

\bibitem{Linde:1982mu}
A.~D. Linde, {\it A new inflationary universe scenario: A possible solution of
  the horizon, flatness, homogeneity, isotropy and primordial monopole
  problems},  {\em Phys. Lett.} {\bf B108} (1982) 389.

\bibitem{Albrecht:1982wi}
A.~Albrecht and P.~J. Steinhardt, {\it Cosmology for grand unified theories
  with radiatively induced symmetry breaking},  {\em Phys. Rev. Lett.} {\bf 48}
  (1982) 1220.

\bibitem{Hartle:1983ai}
J.~B. Hartle and S.~W. Hawking, {\it Wave function of the universe},  {\em
  Phys. Rev.} {\bf D28} (1983) 2960.

\bibitem{Vilenkin:1982de}
A.~Vilenkin, {\it Creation of universes from nothing},  {\em Phys. Lett.} {\bf
  B117} (1982) 25.

\bibitem{Vilenkin:1983xq}
A.~Vilenkin, {\it The birth of inflationary universes},  {\em Phys. Rev.} {\bf
  D27} (1983) 2848.

\bibitem{Tryon:1973}
E.~P. Tryon, {\it Is the universe a vacuum fluctuation ?},  {\em Nature} {\bf
  246} (1973) 396.

\bibitem{Dvali:1998pa}
G.~R. Dvali and S.-H.~H. Tye, {\it Brane inflation},  {\em Phys. Lett.} {\bf
  B450} (1999) 72, [\href{http://arXiv.org/abs/hep-ph/9812483}{{\tt
  hep-ph/9812483}}].

\bibitem{Kachru:2003sx}
S.~Kachru, R.~Kallosh, A.~Linde, J.~Maldacena, L.~McAllister, and S.~P.
  Trivedi, {\it Towards inflation in string theory},  {\em JCAP} {\bf 0310}
  (2003) 013, [\href{http://arXiv.org/abs/hep-th/0308055}{{\tt
  hep-th/0308055}}].

\bibitem{Firouzjahi:2003zy}
H.~Firouzjahi and S.-H.~H. Tye, {\it Closer towards inflation in string
  theory},  {\em Phys. Lett.} {\bf B584} (2004) 147,
  [\href{http://arXiv.org/abs/hep-th/0312020}{{\tt hep-th/0312020}}].

\bibitem{Hsu:2003cy}
J.~P. Hsu, R.~Kallosh, and S.~Prokushkin, {\it On brane inflation with volume
  stabilization},  {\em JCAP} {\bf 0312} (2003) 009,
  [\href{http://arXiv.org/abs/hep-th/0311077}{{\tt hep-th/0311077}}].

\bibitem{Burgess:2004kv}
C.~P. Burgess, J.~M. Cline, H.~Stoica, and F.~Quevedo, {\it Inflation in
  realistic d-brane models},  \href{http://arXiv.org/abs/hep-th/0403119}{{\tt
  hep-th/0403119}}.

\bibitem{Dasgupta:2004dw}
K.~Dasgupta, J.~P. Hsu, R.~Kallosh, A.~Linde, and M.~Zagermann, {\it D3/d7
  brane inflation and semilocal strings},
  \href{http://arXiv.org/abs/hep-th/0405247}{{\tt hep-th/0405247}}.

\bibitem{Gibbons:1978ac}
G.~W. Gibbons, S.~W. Hawking, and M.~J. Perry, {\it Path integrals and the
  indefiniteness of the gravitational action},  {\em Nucl. Phys.} {\bf B138}
  (1978) 141.

\bibitem{Fischler:1990se}
W.~Fischler, D.~Morgan, and J.~Polchinski, {\it Quantum nucleation of false
  vacuum bubbles},  {\em Phys. Rev.} {\bf D41} (1990) 2638.

\bibitem{Halliwell:1989vu}
J.~J. Halliwell and J.~Louko, {\it Steepest descent contours in the path
  integral approach to quantum cosmology. 2. microsuperspace},  {\em Phys.
  Rev.} {\bf D40} (1989) 1868.

\bibitem{Kiefer:1987ft}
C.~Kiefer, {\it Continuous measurement of minisuperspace variables by higher
  multipoles},  {\em Class. Quant. Grav.} {\bf 4} (1987) 1369.

\bibitem{Halliwell:1989vw}
J.~J. Halliwell, {\it Decoherence in quantum cosmology},  {\em Phys. Rev.} {\bf
  D39} (1989) 2912.

\bibitem{Kiefer:1989ud}
C.~Kiefer, {\it Continuous measurement of intrinsic time by fermions},  {\em
  Class. Quant. Grav.} {\bf 6} (1989) 561.

\bibitem{Linde:1984mx}
A.~D. Linde, {\it Quantum creation of the inflationary universe},  {\em Nuovo
  Cim. Lett.} {\bf 39} (1984) 401.

\bibitem{Vilenkin:1985dy}
A.~Vilenkin, {\it Quantum origin of the universe},  {\em Nucl. Phys.} {\bf
  B252} (1985) 141--151.

\bibitem{Steinhardt:1982kg}
P.~J. Steinhardt, {\it Natural inflation},  \href{http://arXiv.org/abs/Very
  Early Universe, Cambridge, England, Jun 21 - Jul 9, 1982}{{\tt Very Early
  Universe, Cambridge, England, Jun 21 - Jul 9, 1982}}.

\bibitem{Linde:1986fd}
A.~D. Linde, {\it Eternally existing selfreproducing chaotic inflationary
  universe},  {\em Phys. Lett.} {\bf B175} (1986) 395.

\bibitem{Dyson:2002pf}
L.~Dyson, M.~Kleban, and L.~Susskind, {\it Disturbing implications of a
  cosmological constant},  {\em JHEP} {\bf 10} (2002) 011,
  [\href{http://arXiv.org/abs/hep-th/0208013}{{\tt hep-th/0208013}}].

\bibitem{Albrecht:2004ke}
A.~Albrecht and L.~Sorbo, {\it Can the universe afford inflation?},
  \href{http://arXiv.org/abs/hep-th/0405270}{{\tt hep-th/0405270}}.

\bibitem{Jones:2002cv}
N.~Jones, H.~Stoica, and S.-H.~H. Tye, {\it Brane interaction as the origin of
  inflation},  {\em JHEP} {\bf 07} (2002) 051,
  [\href{http://arXiv.org/abs/hep-th/0203163}{{\tt hep-th/0203163}}].

\bibitem{Sarangi:2002yt}
S.~Sarangi and S.-H.~H. Tye, {\it Cosmic string production towards the end of
  brane inflation},  {\em Phys. Lett.} {\bf B536} (2002) 185,
  [\href{http://arXiv.org/abs/hep-th/0204074}{{\tt hep-th/0204074}}].

\bibitem{Jones:2003da}
N.~T. Jones, H.~Stoica, and S.-H.~H. Tye, {\it The production, spectrum and
  evolution of cosmic strings in brane inflation},  {\em Phys. Lett.} {\bf
  B563} (2003) 6, [\href{http://arXiv.org/abs/hep-th/0303269}{{\tt
  hep-th/0303269}}].

\bibitem{Pogosian:2003mz}
L.~Pogosian, S.-H.~H. Tye, I.~Wasserman, and M.~Wyman, {\it Observational
  constraints on cosmic string production during brane inflation},  {\em Phys.
  Rev.} {\bf D68} (2003) 023506,
  [\href{http://arXiv.org/abs/hep-th/0304188}{{\tt hep-th/0304188}}].

\bibitem{Copeland:2003bj}
E.~J. Copeland, R.~C. Myers, and J.~Polchinski, {\it Cosmic f- and d-strings},
  \href{http://arXiv.org/abs/hep-th/0312067}{{\tt hep-th/0312067}}.

\bibitem{Dvali:2003zj}
G.~Dvali and A.~Vilenkin, {\it Formation and evolution of cosmic d-strings},
  {\em JCAP} {\bf 0403} (2004) 010,
  [\href{http://arXiv.org/abs/hep-th/0312007}{{\tt hep-th/0312007}}].

\bibitem{Leblond:2004uc}
L.~Leblond and S.-H.~H. Tye, {\it Stability of d1-strings inside a d3-brane},
  {\em JHEP} {\bf 03} (2004) 055,
  [\href{http://arXiv.org/abs/hep-th/0402072}{{\tt hep-th/0402072}}].

\bibitem{Jackson:2004zg}
M.~G. Jackson, N.~T. Jones, and J.~Polchinski, {\it Collisions of cosmic f- and
  d-strings},  \href{http://arXiv.org/abs/hep-th/0405229}{{\tt
  hep-th/0405229}}.


\bibitem{Chamseddine:1992qu}
A.~H. Chamseddine, {\it A Study of noncritical strings in arbitrary dimensions},
{\em Nucl. Phys.} {\bf B368} (1992) 98.

\bibitem{Silverstein:2001xn}
E.~Silverstein, {\it (A)dS backgrounds from asymmetric orientifolds},
 \href{http://arXiv.org/abs/hep-th/0106209}{{\tt  hep-th/0106209}}.

\bibitem{Maloney:2002rr}
A.~Maloney, E.~Silverstein and A.~Strominger,
{\it De Sitter space in noncritical string theory},
\href{http://arXiv.org/abs/hep-th/0205316}{{\tt  hep-th/0205316}}.

\bibitem{Kofman:2004yc}
L.~Kofman {\em et.~al.}, {\it Beauty is attractive: Moduli trapping at enhanced
  symmetry points},  \href{http://arXiv.org/abs/hep-th/0403001}{{\tt
  hep-th/0403001}}.

\bibitem{Feng:2000if}
J.~L. Feng, J.~March-Russell, S.~Sethi, and F.~Wilczek, {\it Saltatory
  relaxation of the cosmological constant},  {\em Nucl. Phys.} {\bf B602}
  (2001) 307, [\href{http://arXiv.org/abs/hep-th/0005276}{{\tt
  hep-th/0005276}}].

\bibitem{Bousso:2000xa}
R.~Bousso and J.~Polchinski, {\it Quantization of four-form fluxes and
  dynamical neutralization of the cosmological constant},  {\em JHEP} {\bf 06}
  (2000) 006, [\href{http://arXiv.org/abs/hep-th/0004134}{{\tt
  hep-th/0004134}}].

\bibitem{Hawking:1980gf}
S.~W. Hawking, {\it The path integral approach to quantum gravity},
  \href{http://arXiv.org/abs/General Relativity: An Einstein Centenary Survey
  (Cambridge U. Press, 1979)}{{\tt General Relativity: An Einstein Centenary
  Survey (Cambridge U. Press, 1979)}}.

\bibitem{Gutperle:2002ai}
M.~Gutperle and A.~Strominger, {\it Spacelike branes},  {\em JHEP} {\bf 04}
  (2002) 018, [\href{http://arXiv.org/abs/hep-th/0202210}{{\tt
  hep-th/0202210}}].

\bibitem{Zurek:1982ii}
W.~H. Zurek, {\it Environment induced superselection rules},  {\em Phys. Rev.}
  {\bf D26} (1982) 1862.

\bibitem{Zeh:1988ws}
H.~D. Zeh, {\it Time in quantum gravity},  {\em Phys. Lett.} {\bf A126} (1988)
  311.

\bibitem{Mellor:1991mx}
F.~Mellor, {\it Decoherence in quantum kaluza-klein theories},  {\em Nucl.
  Phys.} {\bf B353} (1991) 291--301.

\bibitem{Kiefer:1992cn}
C.~Kiefer, {\it Decoherence in quantum electrodynamics and quantum gravity},
  {\em Phys. Rev.} {\bf D46} (1992) 1658--1670.

\bibitem{Hawking:2002af}
S.~W. Hawking and T.~Hertog, {\it Why does inflation start at the top of the
  hill?},  {\em Phys. Rev.} {\bf D66} (2002) 123509,
  [\href{http://arXiv.org/abs/hep-th/0204212}{{\tt hep-th/0204212}}].

\bibitem{Lu:2004fe}
H.~Lu, J.~F. Vazquez-Poritz, and J.~E. Wang, {\it De sitter bounces},
  \href{http://arXiv.org/abs/hep-th/0406028}{{\tt hep-th/0406028}}.

\bibitem{Bousso:2002fq}
R.~Bousso, {\it Adventures in de sitter space},
  \href{http://arXiv.org/abs/hep-th/0205177}{{\tt hep-th/0205177}}.

\bibitem{Brandenberger:1989aj}
R.~H. Brandenberger and C.~Vafa, {\it Superstrings in the early universe},
  {\em Nucl. Phys.} {\bf B316} (1989) 391.

\bibitem{Alexander:2000xv}
S.~Alexander, R.~H. Brandenberger, and D.~Easson, {\it Brane gases in the early
  universe},  {\em Phys. Rev.} {\bf D62} (2000) 103509,
  [\href{http://arXiv.org/abs/hep-th/0005212}{{\tt hep-th/0005212}}].

\bibitem{Watson:2003gf}
S.~Watson and R.~Brandenberger, {\it Stabilization of extra dimensions at tree
  level},  {\em JCAP} {\bf 0311} (2003) 008,
  [\href{http://arXiv.org/abs/hep-th/0307044}{{\tt hep-th/0307044}}].

\bibitem{Easther:2002mi}
R.~Easther, B.~R. Greene, and M.~G. Jackson, {\it Cosmological string gas on
  orbifolds},  {\em Phys. Rev.} {\bf D66} (2002) 023502,
  [\href{http://arXiv.org/abs/hep-th/0204099}{{\tt hep-th/0204099}}].

\bibitem{Easther:2002qk}
R.~Easther, B.~R. Greene, M.~G. Jackson, and D.~Kabat, {\it Brane gas cosmology
  in m-theory: Late time behavior},  {\em Phys. Rev.} {\bf D67} (2003) 123501,
  [\href{http://arXiv.org/abs/hep-th/0211124}{{\tt hep-th/0211124}}].

\bibitem{Burgess:2001fx}
C.~P. Burgess, M.~Majumdar, D.~Nolte, F.~Quevedo, G.~Rajesh, and R.~Zhang, {\it
  The inflationary brane-antibrane universe},  {\em JHEP} {\bf 07} (2001) 047,
  [\href{http://arXiv.org/abs/hep-th/0105204}{{\tt hep-th/0105204}}].

\bibitem{Alexander:2001ks}
S.~H.~S. Alexander, {\it Inflation from d - anti-d brane annihilation},  {\em
  Phys. Rev.} {\bf D65} (2002) 023507,
  [\href{http://arXiv.org/abs/hep-th/0105032}{{\tt hep-th/0105032}}].

\bibitem{Dvali:2001fw}
G.~R. Dvali, Q.~Shafi, and S.~Solganik, {\it D-brane inflation},
  \href{http://arXiv.org/abs/hep-th/0105203}{{\tt hep-th/0105203}}.

\bibitem{Buchan:2003gx}
S.~Buchan, B.~Shlaer, H.~Stoica, and S.-H.~H. Tye, {\it Inter-brane
  interactions in compact spaces and brane inflation},  {\em JCAP} {\bf 0402}
  (2004) 013, [\href{http://arXiv.org/abs/hep-th/0311207}{{\tt
  hep-th/0311207}}].

\bibitem{Klebanov:2000hb}
I.~R. Klebanov and M.~J. Strassler, {\it Supergravity and a confining gauge
  theory: Duality cascades and chisb-resolution of naked singularities},  {\em
  JHEP} {\bf 08} (2000) 052, [\href{http://arXiv.org/abs/hep-th/0007191}{{\tt
  hep-th/0007191}}].

\bibitem{Garriga:1999bq}
J.~Garriga and M.~Sasaki, {\it Brane-world creation and black holes},  {\em
  Phys. Rev.} {\bf D62} (2000) 043523,
  [\href{http://arXiv.org/abs/hep-th/9912118}{{\tt hep-th/9912118}}].

\bibitem{Hawking:2000kj}
S.~W. Hawking, T.~Hertog, and H.~S. Reall, {\it Brane new world},  {\em Phys.
  Rev.} {\bf D62} (2000) 043501,
  [\href{http://arXiv.org/abs/hep-th/0003052}{{\tt hep-th/0003052}}].

\bibitem{Dewitt:1967yk}
B.~S. Dewitt, {\it Quantum theory of gravity. 1. the canonical theory},  {\em
  Phys. Rev.} {\bf 160} (1967) 1113--1148.

\bibitem{Vilenkin:1986cy}
A.~Vilenkin, {\it Boundary conditions in quantum cosmology},  {\em Phys. Rev.}
  {\bf D33} (1986) 3560.

\bibitem{Rubakov:1984bh}
V.~A. Rubakov, {\it Quantum mechanics in the tunneling universe},  {\em Phys.
  Lett.} {\bf B148} (1984) 280.

\bibitem{Chiu:1977ds}
C.~B.~Chiu, E.~C.~G.~Sudarshan and B.~Misra,
{\it Time Evolution Of Unstable Quantum States And A Resolution Of Zeno's
Paradox},
{\em Phys. Rev. D} {\bf 16} (1977) 520.

\bibitem{Peres:1980ux}
A.~Peres,
{\it Zeno Paradox In Quantum Theory},
{\em Am. J. Phys.}  {\bf 48}  (1980) 931.

\bibitem{Halliwell:1985eu}
J.~J. Halliwell and S.~W. Hawking, {\it The origin of structure in the
  universe},  {\em Phys. Rev.} {\bf D31} (1985) 1777.

\bibitem{Horowitz:2003he}
G.~T. Horowitz and J.~Maldacena, {\it The black hole final state},  {\em JHEP}
  {\bf 02} (2004) 008, [\href{http://arXiv.org/abs/hep-th/0310281}{{\tt
  hep-th/0310281}}].

\end{thebibliography}
\providecommand{\href}[2]{#2}\begingroup\raggedright\endgroup

\end{document}